\documentclass[pre,superscriptaddress,twocolumn]{revtex4-1}
\pdfoutput=1
\usepackage[colorlinks=true,urlcolor=blue]{hyperref}
\usepackage{amsfonts}
\usepackage{graphicx}
\usepackage{mathrsfs}
\usepackage{amsmath}
\usepackage{amssymb}
\usepackage{xcolor}
\usepackage{bbold}
\usepackage{bm}

\usepackage{color}
\definecolor{red}{rgb}{0.75,0,0}
\definecolor{blue}{rgb}{0,0,0.75}
\definecolor{green}{rgb}{0,0.5,0}
\definecolor{black}{rgb}{0,0,0}

\newcommand{\red}[1]{{\color{black} #1}}

\DeclareMathOperator{\tr}{tr}

\begin{document}

\title{Confinement-induced self-organization in growing bacterial colonies}
 
\author{Zhihong You}
\affiliation{Department of Physics, University of California Santa Barbara, Santa Barbara, CA 93106, USA}
\affiliation{Instituut-Lorentz, Universiteit Leiden, P.O. Box 9506, 2300 RA Leiden, The Netherlands}
\author{Daniel J. G. Pearce}
\affiliation{Department of Theoretical Physics, Universit\'e de Gen\`eve, 1205 Gen\`eve, Switzerland}
\author{Luca Giomi}
\email[Corresponding author:\ ]{giomi@lorentz.leidenuniv.nl}
\affiliation{Instituut-Lorentz, Universiteit Leiden, P.O. Box 9506, 2300 RA Leiden, The Netherlands}

\begin{abstract}
\vspace*{0.25cm}
We investigate the emergence of global alignment in colonies of dividing rod-shaped cells under confinement. Using molecular dynamics simulations and continuous modeling, we demonstrate that geometrical anisotropies in the confining environment give rise to imbalance in the normal stresses, which, in turn, drives a collective rearrangement of the cells. This behavior crucially relies on the colony's solid-like mechanical response at short time scales and can be recovered within the framework of active hydrodynamics upon modeling bacterial colonies as growing viscoelastic gels characterized by Maxwell-like stress relaxation.  
\vspace*{1.5cm}
\end{abstract}

\maketitle

\section{Introduction}

The ability {to exploit} the physical properties of the environment, in order to achieve biological functionality at the large scale, is the hallmark of self-organization in multicellular systems \cite{Friedl:2004,Haeger:2015,BenJacob:1998,BenJacob:2000}. Both in eukaryotes \cite{Friedl:2004,Haeger:2015} and prokaryotes \cite{BenJacob:1998,BenJacob:2000}, cells can take advantage of the geometrical and mechanochemical features of their surrounding, such as confinement, friction, compliance and degradability, in order to migrate and proliferate. In prokaryotes, a spectacular example of this behavior is offered by growing colonies of sessile bacteria. Although lacking of a self-propulsion machinery, these cells are still capable of taking advantage of the extensile forces resulting from their individual growth to achieve collective migration, while relying on the environmental topography to navigate \cite{Allen:2018}.

Volfson {\em et al}., investigated this mechanism using a monolayer of dividing non-motile {\em E. coli} cells cultured in microfluidic channel with open ends, that allow the cells to escape \cite{Volfson:2008}. Such a confined bacterial layer is initially structureless, but, as the density progressively increases, global nematic order starts to develop within the system, with the majority of the cells oriented along the longitudinal direction of the channel. This alignment mechanism, resulting from the combination of growth and confinement, allows the cells to efficiently {relieve} the internal extensile stresses and prevents the colony from overcrowding by facilitating the emergence of an expansion flow directed toward the channel outlets. In long channels, however, such a globally ordered state is unstable and domains of non-longitudinally oriented cells arise throughout the colony, rendering it disordered at the large scale \cite{Boyer:2011}. 

Whereas the latter instability, triggering the transition from global alignment to disorder, is well understood, the nature of the mechanism leading to global alignment is still debated, despite dramatically affecting the overall fitness of the colony. Volfson \textit{et al}. proposed that global longitudinal alignment, arises in response to the bacterial expansion flow, thanks to the propensity of nematic liquid crystals to align with respect to the flow direction. Thus, the channel geometry dictates the direction of the expansion flow, which, in turn, acts as an ordering field for the bacteria themselves. A different mechanism, recently proposed by Karamched {\em et al}., revolves instead around the role of parallel anchoring at the channel walls \cite{Karamched:2019}. Using a stochastic model based on the spatial Moran process, they demonstrate that global alignment can emerge even in the absence of flow-alignment. The latter finding is, however, built upon specifically designed growth dynamics, which, based on previous experiments and simulations \cite{Volfson:2008, Cho:2007, Sheats:2017, Boyer:2011, Fuentes:2013}, are by no means indispensable for the occurrence of global alignment {\em in vitro}.

In this article we propose an alternative explanation, rooted in the emergence of globally anisotropic stresses in the confined bacterial population. Using computer simulations of a hard-rod model, we find that cell growth gives rise to a persistent accumulation of mechanical stress in the colony. While longitudinal stresses can be efficiently relieved via the expansion flow toward the channel outlets, lateral confinement prohibits cell motion in the transverse direction, leading to a build up of transversal stress. The resulting mechanical anisotropy determines a net torque that reorients the cells in the direction of minimal stress, which, in the case of a straight channel, coincides with the longitudinal direction. This interpretation is additionally supported by a continuum model of the expanding colony as an active viscoelastic nematic gel, characterized by Maxwell-like stress relaxation. These discoveries not only deepen our understanding on the roles of confinement and mechanical stresses on the self-organization of growing bacterial colonies, providing a potential way to control growing bacterial colonies, but also sheds new light on the interplay between orientation and stress in active viscoelastic liquid crystals.

\section{\label{sec:results}Results and discussion}
\subsection{\label{sec:alignment}\red{Growing colonies under confinement}}

\begin{figure*}[t]
\centering
\includegraphics[width=1.0\textwidth]{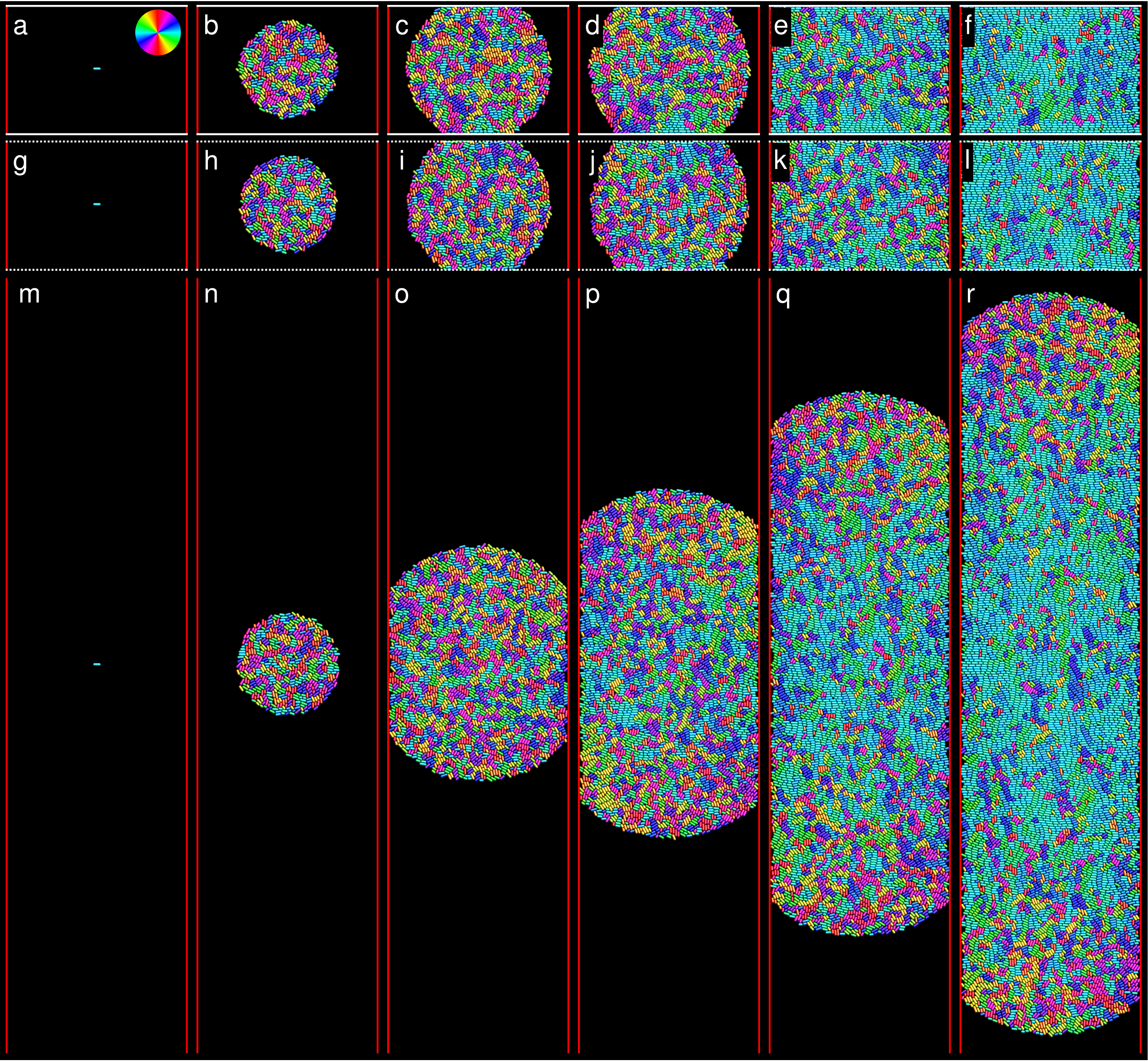}
\caption{\label{fig:snapshots-grow} \red{\textbf{Colonization under confinement.} Snapshots of growing colonies at different time points. The boundary conditions are (a--f) rigid wall confinement, (g--l) periodic confinement, and (m--r) no confinement in the $y$ direction.} Cells are color-coded by their orientation, as indicated by the color wheel in panel (a). In all three colonies, $L_{x}=70\,\mu$m, while $L_{y}=50\,\mu$m in (a--l). The colony height in (r) is about $300\,\mu$m. \red{In all three scenarios the direction of the confining channel is orthogonal to the red outlets, hence horizontal.}}
\end{figure*}

We use a minimal model of proliferating hard rods to study growing bacteria under confinement \cite{Farrell:2013, Grant:2014, You:2018}. \red{Each cell is modeled as a spherocylinder, which can grow, divide, and push away its neighbors as it elongates. Details of the model can be found in Sec. \ref{sec:hdmodel}.} Figs.~\ref{fig:snapshots-grow}a--\ref{fig:snapshots-grow}r illustrate the typical dynamics of our {\em in silico} colonies, subject to boundary conditions of three different types. When the colony is confined by rigid walls (Figs.~\ref{fig:snapshots-grow}a--\ref{fig:snapshots-grow}f), our results reproduce previous experiments and simulations \cite{Volfson:2008, Cho:2007, Sheats:2017, Boyer:2011, Fuentes:2013, Winkle:2017, Karamched:2019, Alnahhas:2019}: the long time dynamics consists of a steady state characterized by global nematic order, with the majority of the cells aligned along the horizontal direction. A possible explanation of this phenomenon, proposed in Ref. \cite{Karamched:2019}, relies on the intuitive idea that steric repulsion from the rigid walls generate torques on the peripheral cells, thereby aligning these horizontally. Such an alignment then propagates to the interior of the channel as a consequence of the cell-cell steric repulsion, resulting into uniform horizontal alignment across the entire colony. The latter mechanism occurs in thermotropic nematic liquid crystals (see e.g. Ref. \cite{DeGennes:1993}) and was recently observed in confined monolayers of eukaryotic spindle-like cells \cite{Duclos:2014}. However, our results on periodically confined colony (Figs.~\ref{fig:snapshots-grow}g--\ref{fig:snapshots-grow}l) readily disprove this hypothesis. In this instance the colony is not confined by rigid horizontal walls, however global horizontal alignment is still prominent. Specifically, once the top and bottom boundaries \red{of the colony} merge, disconnected domains of horizontally aligned cells start to emerge uniformly throughout the colony (Fig.~\ref{fig:snapshots-grow}j). Over time, these domains expand and eventually form a percolating cluster spanning the entire length of the colony (Figs.~\ref{fig:snapshots-grow}k and \ref{fig:snapshots-grow}l). Remarkably, global alignment emerges even when vertical confinement is removed (Figs.~\ref{fig:snapshots-grow}m--\ref{fig:snapshots-grow}r). Here, a region of horizontally aligned cells appears in the center of the channel once the colony height exceeds a given threshold and then gradually expands vertically, leaving two disordered caps of constant thickness at the top and bottom of the colony. Finally, increasing the channel width eventually disrupts global horizontal alignment (see \red{Fig. S1 in the} Supplementary Information). \red{In particular, the wider the channel, the larger the height of the colony at which the horizontally aligned region first appears. In addition, nematic order in wide channels is lower on average and the disordered caps are thicker than those in narrow channels. The latter phenomenon is well suited to be tested with {\em in vitro} experiments, for instance by growing multiple bacterial colonies within confining channels of varying width.} 

\subsection{\label{sec:hdmodel}Anisotropic stress drives cell reorientation}

Volfson {\em at al}. \cite{Volfson:2008} proposed that the longitudinal alignment of {\em E. coli} bacteria in a channel, resulted from the combination of the expansion flow, caused by cell division, and the tendency of nematic liquid crystals to reorient in the presence of a velocity gradient. Whereas plausible to explain the observed longitudinal alignment within a straight channel, however, this mechanism would also determine preferential radial alignment in freely expanding colonies, as a consequence of prominent radial flow characterizing these systems (see e.g. Ref. \cite{You:2018}). By contrast, such a behavior has never been observed in experiments and simulations \cite{DellArciprete:2018}. Furthermore, longitudinal alignment can also emerge in the absence of flow, provided the cells are subject to anisotropic normal stresses. 

\begin{figure}
\centering
\includegraphics[width=1.0\linewidth]{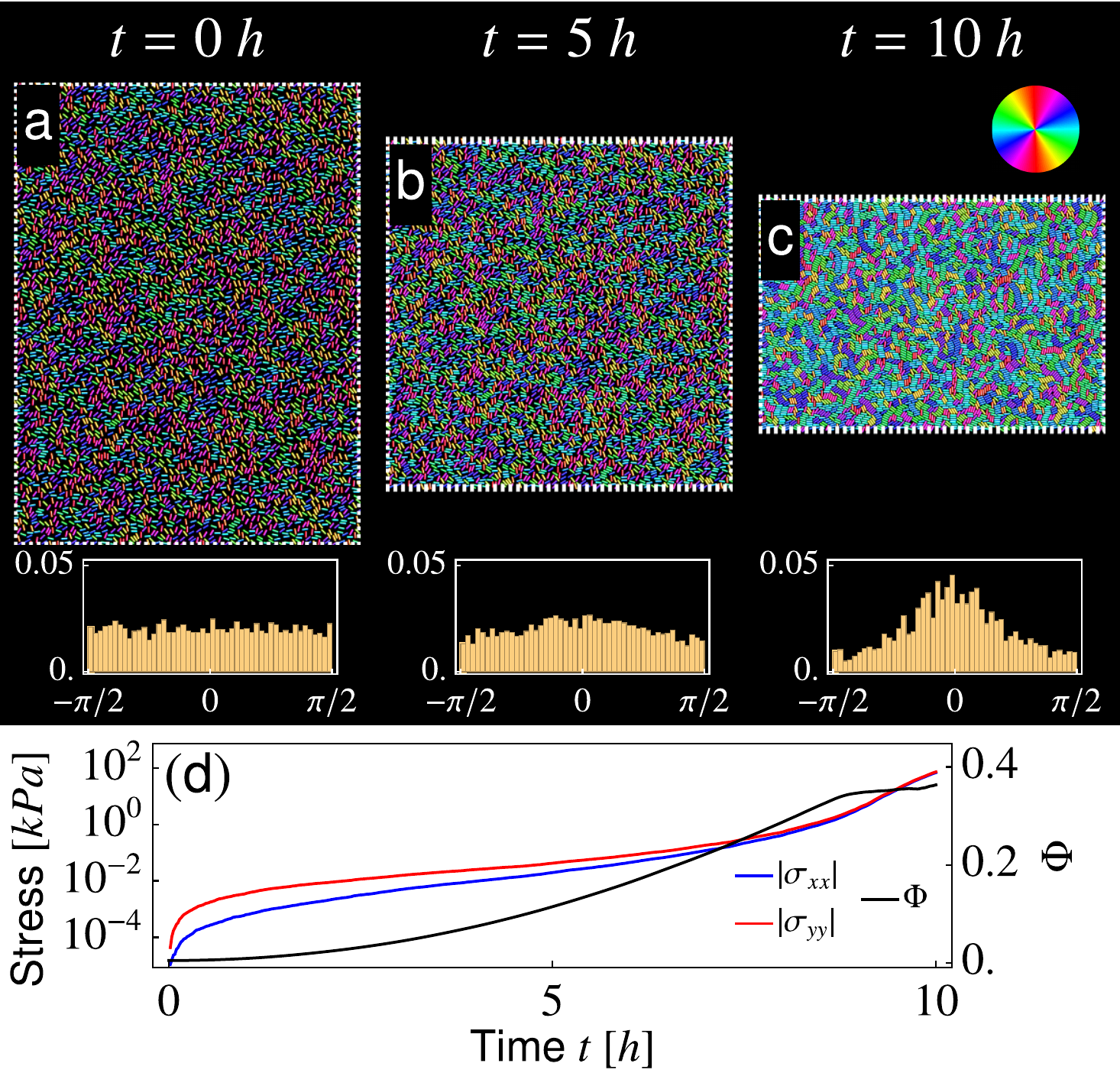}
\caption{\label{fig:shrink_box} \textbf{Dynamics of the shrinking colony.} (a--c) Snapshots at different time points. Cells are color-coded by their orientations according to the color wheel in panel (c). The insets show the histograms of cell orientation at the corresponding time points. (d) The average stresses and the alignment parameter $\Phi$ as functions of \red{time}. The system contains $5000$ cells in total, with a width $L_{x}=150\,\mu$m and an initial height $L_{y}=200\,\mu$m. The shrinking speed, i.e. the relative speed between the top and the bottom boundaries, is $V_{y}=10\,\mu$m/h.}
\end{figure}

To demonstrate this latter \red{statement,} we have simulated an initially disordered distribution of non-growing cells, confined in a rectangular doubly periodic box and subject to a uniform contraction along the $y-$direction (Fig.~\ref{fig:shrink_box}). Initially, each cell has a random position and orientation (Fig.~\ref{fig:shrink_box}a). Then, the box height is uniformly shortened by mean of the scaling transformation: $y(t+\Delta t)=k'y(t)$, where $y$ is the abscissa of the cells and the horizontal boundaries, $k'=(L_{y}-V_{y}\Delta t)/L_{y}$ the shrinking factor and $V_{y}$ the relative speed of the two horizontal boundaries. Note that cell orientations remain unchanged during the scaling. During the process, we measure the stresses experienced by the cells using the virial construction, namely:
\begin{equation}\label{eq:virial_stress}
\bm{\sigma}^{i}=\frac{1}{a'_{i}}\sum_{j}  \bm{r}_{ij}\,\bm{F}_{ij}^{c}\;,
\end{equation}
where $a'_i=a_i/\phi$ is the effective area occupied by the $i-$th cell, with \red{$a_{i}=d_{0}(l_{i}+\pi d_{0}/4)$ the exact cell area}, $\phi$ the local packing fraction and $\bm{r}_{ij}$ is the position of point of contact between the $i-$th and the $j-$th cell with respect to the center of mass of the $i-$th cell. To characterize the amount of global orientation, we define the following alignment parameter:
\begin{equation}\label{eq:phi}
\Phi = 2\langle p_{x}^{2}\rangle -1\;,
\end{equation}
where $\langle \cdot \rangle$ denotes an average over all cells in the box. By construction $-1 \le \Phi \le 1$, with $\Phi=0$ corresponding to a completely disordered configuration and $\Phi=\pm 1$ to perfect alignment in $x$ and $y$, respectively. Figs.~\ref{fig:shrink_box}a--\ref{fig:shrink_box}c summarize the dynamics of the shrinking colony. The system is initially isotropic and the orientation of the cells is the uniformly distributed (Fig.~\ref{fig:shrink_box}a and its inset). Shrinking increases the cells' overlap, hence the stress across the colony. However, because of the anisotropy in the scaling transformation, the vertical normal stress $\sigma_{yy}$ increases more quickly than the horizontal normal stress $\sigma_{xx}$ (Fig.~\ref{fig:shrink_box}d). Simultaneously, longitudinal alignment develops throughout the colony (\red{Fig.~\ref{fig:shrink_box}c--\ref{fig:shrink_box}d}). 

\begin{figure}[t]
\centering
\includegraphics[width=1.0\linewidth]{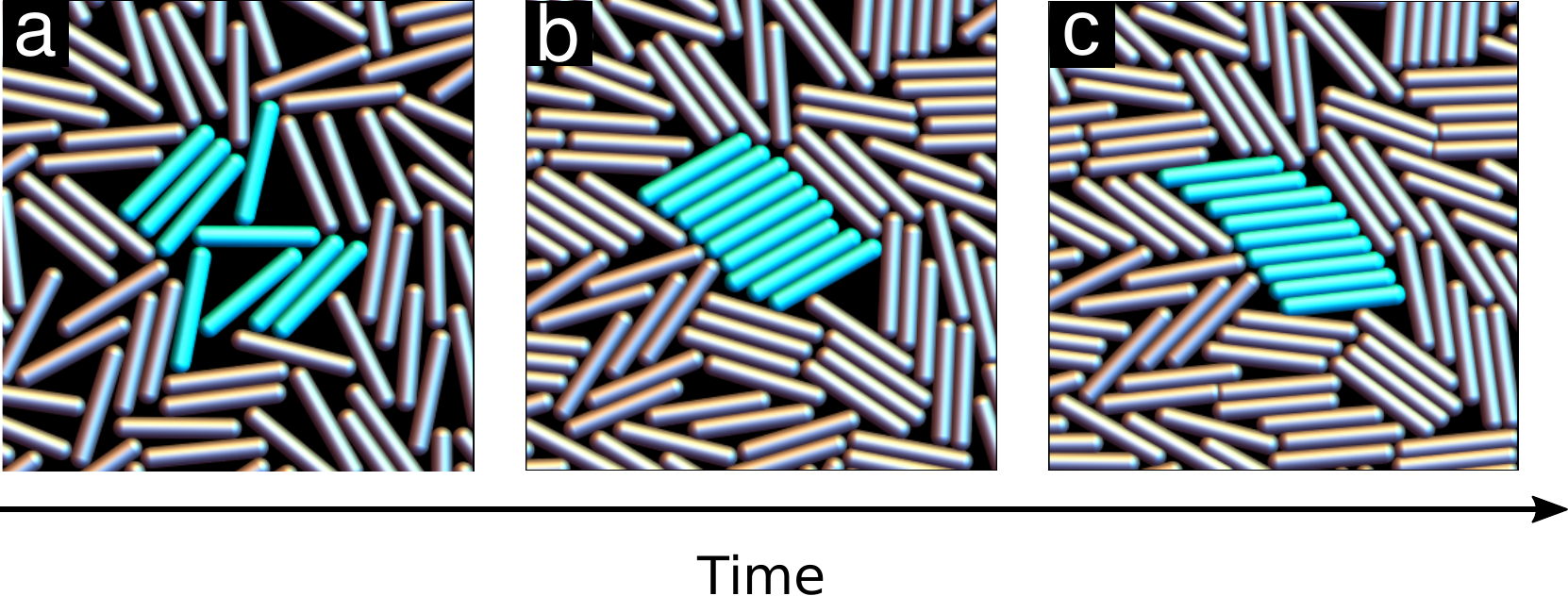}
\caption{\label{fig:domain_torque} \textbf{Alignment mechanism in uniformly shrunk colonies.} (a) The cells in the central part the panel are initially loosely packed and do not exhibit a preferential orientation. (b) As time progresses, the shrinkage causes them to form a stack of nine tightly packed cells roughly oriented at $45^{\circ}$ with respect to the horizontal direction. (c) Finally, as $\sigma_{yy}>\sigma_{xx}$, this cluster undergoes a collective rearrangement, whose effect is to redistribute the stress among the normal components by reorienting the cells along the horizontal direction. \red{Snapshots corresponds to different times in the simulations of a shrunk colony.} }
\end{figure}

These observations suggest a causal relation between the occurrence of transient anisotropy in the normal stresses and the emergence of longitudinal alignment within the bacterial population. Here we postulate the following mechanism. Passive spherocylinders in proximity of the isotropic-nematic phase transition are known to organize in clusters consisting of highly aligned cells, sometimes referred to as cybotactic clusters \cite{DeVries:1970,Frenkel:1987}. This effect is further enhanced in colonies of growing bacteria, as a consequence of the long-wavelength instability of the nematic ground state driven by the extensile active stresses \cite{You:2018}. These clusters are not held together by attractive interactions and eventually break up or merge into larger domains. At short time scales, however, clusters may exhibit solid-like mechanical response to environmental forces and, in particular, undergo internal rearrangements while subject to anisotropic stresses, resulting in a redistribution of the stress that eases the local \red{stress} anisotropy. Fig.~\ref{fig:domain_torque} illustrates this process in the case of uniformly shrunk colonies. The cells in the central part of Fig.~\ref{fig:domain_torque}a are initially loosely packed and do not exhibit a preferential orientation, but, as time progresses, the shrinkage causes them to form a stack of nine tightly packed cells roughly oriented at $45^{\circ}$ with respect to the horizontal direction (Fig.~\ref{fig:domain_torque}b). Finally, as $\sigma_{yy}>\sigma_{xx}$, this cluster undergoes a collective rearrangement, whose effect is to redistribute the stress among the normal components by reorienting the cells along the horizontal direction (Fig.~\ref{fig:domain_torque}c).   

Some remarks are in order. First, the aligning mechanism illustrated above in the case of uniformly shrunk colonies results from the orchestrated action of multiple effects. These include the entropic alignment of the cells originating from their excluded volume interactions, further enhanced by the progressive increase in density, and the \red{cluster's} collective rotation in the presence of anisotropic normal stresses. The latter process, in turn, crucially relies on the fact that cells can slide along their longitudinal direction $\bm{p}$, while their motion in the transversal direction $\bm{p}^{\perp}$ is obstructed. Second, because of the absence of attractive interactions among our {\em in silico} cells, the same effect cannot be produced solely by shear stresses, as these would mainly slide the cells with respect to each other, resulting into a spreading of the domain along the horizontal direction. Real bacteria do exhibit attractive interactions, mediated by proteins such as the adhesin FimH \cite{Sauer:2016}, but these are inevitably weaker than the repulsive excluded volume interactions among cells. Finally, such an alignment mechanism evidently relies on the viscoelastic nature of bacterial layers and its performance depends on how the time required for the stress anisotropic to build up compares with the viscoelastic crossover time. In Sec.~\ref{sec:continuum}, we directly test this mechanism by implementing it in a continuum viscoelastic model of growing colonies characterized by Maxwell-like stress relaxation.

\begin{figure}[t]
\centering
\includegraphics[width=1.0\linewidth]{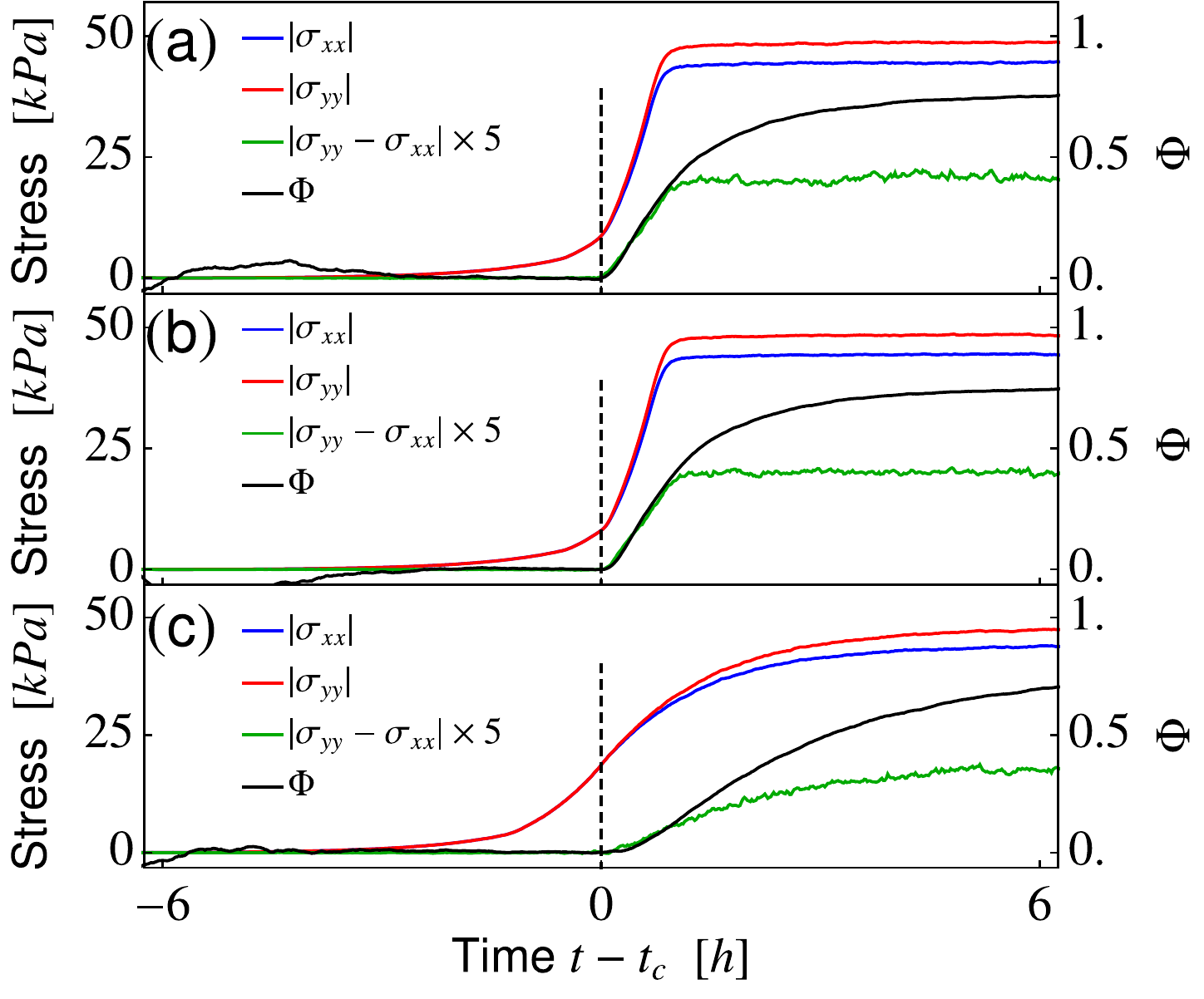}
\caption{\label{fig:stre-cent} \red{\textbf{Stress dynamics in growing colonies.} Normal stresses and the alignment parameter in region $\mathcal{R}_{0}$ of the colonies. The boundary conditions are} (a) rigid confinement \red{(see e.g. Figs. \ref{fig:snapshots-grow}a--\ref{fig:snapshots-grow}f)}, (b) periodic confinement \red{(Figs. \ref{fig:snapshots-grow}g--\ref{fig:snapshots-grow}l)}, and (c) no confinement \red{(Figs. \ref{fig:snapshots-grow}m--\ref{fig:snapshots-grow}r)}. \red{In (a)--(b)}, the black dashed line indicates $t_{c}$, the time at which the colonies reach the \red{horizontal} boundary. \red{In panel (c), $t_{c}$ is set as the time where the normal stress difference in region $\mathcal{R}_{0}$ becomes nonzero. The results at each boundary condition are obtained by averaging over $200$ simulations of growing colonies.}}
\end{figure}

\subsection{Stress anisotropy and alignment in growing colonies}

To test whether the mechanism proposed in the previous section carries over to colonies of growing bacteria, we look back at the expanding colonies depicted in Fig.~\ref{fig:snapshots-grow} and track the alignment parameter $\Phi$ as well as the normal stresses $\sigma_{xx}$ and $\sigma_{yy}$ during expansion. Generally speaking, the normal stress decreases with the distance from the colony center and increases monotonically in time. To capture such spatial-temporal dynamics more precisely, let us focus on a region $\mathcal{R}_{0}$, defined as the area within $20\,\mu$m from the colony center: i.e. $\sqrt{x^{2}+y^{2}}<20\,\mu$m. Figs.~\ref{fig:stre-cent}a--\ref{fig:stre-cent}c show the average normal stresses and the alignment parameter within $\mathcal{R}_{0}$ of the three colonies displayed in Fig.~\ref{fig:snapshots-grow}, subject to different boundary conditions. As the fronts of the colonies approach the boundaries, stress is globally isotropic and the colonies have the typical structure observed in free space \cite{You:2018}. This behavior drastically changes once the colonies meet the boundaries at $t=t_{c}$. In particular, in colonies subject to hard-wall (Figs.~\ref{fig:snapshots-grow}a-f) or periodic confinement (Figs.~\ref{fig:snapshots-grow}g-l), both normal stresses undergo a dramatic increase, hence $\sigma_{yy}>\sigma_{xx}$ as a consequence of the inhibition of the bacterial motion in the $y-$direction (Figs.~\ref{fig:stre-cent}a--\ref{fig:stre-cent}b). Remarkably, a similar behavior is also found in unconfined colonies (Figs.~\ref{fig:snapshots-grow}m-r). Here stress builds up more gently than in the presence of hard walls and yet $\sigma_{yy}>\sigma_{xx}$ at any time $t>t_{c}$ (Fig.~\ref{fig:stre-cent}c). This results from the fact that, while the removal of cells in the $x-$direction relieves the associated normal stress $\sigma_{xx}$, $\sigma_{yy}$ increases with time as the colony elongates along the $y-$direction. Intuitively, such an {\em implicit} confinement is much weaker than the {\em explicit} confinement provided by a hard wall or a periodic boundary, thus stresses increase more gradually. In all three colonies, the occurrence of stress anisotropy, represented by the non-vanishing normal stress difference $|\sigma_{yy}-\sigma_{xx}|$, continuously drives cells to align with the horizontal direction. Eventually, a steady state is reached where the majority of the cells are aligned along the $x-$direction, thus $\Phi\gtrsim 0.5$, and $\sigma_{yy}\gtrsim \sigma_{xx}$ (see e.g. Figs.~\ref{fig:snapshots-grow}f,l and r). 

Despite the aligning mechanisms having the same origin in both shrinking and growing colonies, the latter exhibit a number of system-specific properties that are not found in passive systems. First, shrinking colonies undergo a jamming transition for sufficiently high densities; this effectively freezes the cells' orientation. \red{By contrast, in growing colonies, the perpetual cell growth and division prevent the system from jamming, thus allowing the colony to further improve its global longitudinal alignment. This is exemplified by the fact that $\Phi$ is systematically larger in growing colonies than in shrinking ones.} Second, because of the aligning mechanism described in the previous section, domains of horizontally aligned cells are more stable than others. As cells duplicate, these horizontal domains can then expand without being reoriented and eventually take over the entire colony. Finally, as we detail in Sec.~\ref{sec:continuum}, cell growth results in a continuous injection of {\em active} stress $\bm{\sigma}^{\rm a} \sim \langle \bm{p}\bm{p} \rangle$. This gives rise to a feedback mechanism that directly converts orientational anisotropy into stress anisotropy, in such a way as to enhance the redistribution of stress from the transverse to the longitudinal component. This effect stabilizes the horizontally aligned configuration, where the normal stress difference $|\sigma_{yy}-\sigma_{xx}|$ is minimal at a given cell density.      

\begin{figure}[t]
\centering
\includegraphics[width=1.0\linewidth]{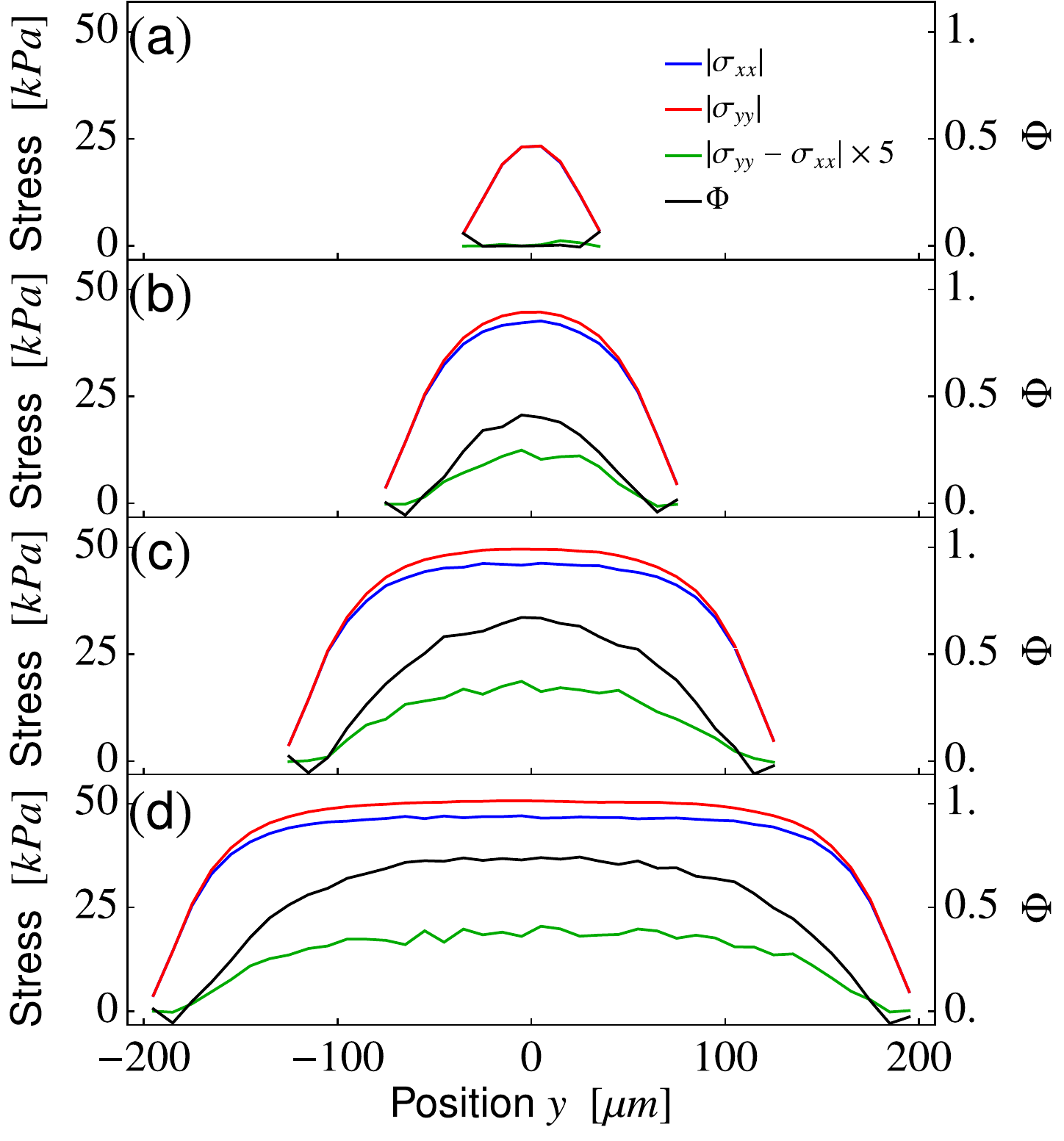}
\caption{\label{fig:StreA-y} \textbf{Stress distribution in growing colonies.} Spatial distribution of stresses and alignment parameter $\Phi$ in a $20\,\mu$m width region located at the center of the colony and spanning its entire height at different times. All the quantities are measured upon dividing the region into smaller boxes of height $10\,\mu$m and calculating their averages within each box. The channel is $L_{x}=70\,\mu$m \red{wide}, whereas the colony's height is: (a) 70 $\mu$m, (b) 150 $\mu$m, (c) 250 $\mu$m, (d) 390 $\mu$m. \red{The results are obtained by averaging over $100$ simulations of growing colonies.}}
\end{figure}

To further check our hypothesis, we explore the onset of stress anisotropy and longitudinal alignment in the unconfined colony displayed in Figs.~\ref{fig:snapshots-grow}o--\ref{fig:snapshots-grow}r. For this purpose, we focus on a $20\,\mu$m width region located at the center of the colony and spanning its entire height. Fig.~\ref{fig:StreA-y} shows the spatial distribution of the stresses and the alignment parameter $\Phi$ at different times, or, equivalently, colony heights. At \red{early} times and for small colony heights, stress is isotropic and the cells have random orientation (Fig.~\ref{fig:StreA-y}a). Once the colony becomes sufficiently elongated in the $y-$direction, however, the normal stresses in the center of the colony start to \red{differentiate} as a consequence of the effective confinement originating from the top and bottom caps (Fig.~\ref{fig:StreA-y}b). The anisotropy increases with time as the colony progressively elongates (Fig.~\ref{fig:StreA-y}c--Fig.~\ref{fig:StreA-y}d). Importantly, the alignment parameter $\Phi$ and the normal stress difference $|\sigma_{yy}-\sigma_{xx}|$ transition from vanishing to positive simultaneously and continue evolving in parallel at any later time, thus confirming the correlation between these two quantities.  

\begin{figure}[t]
\centering
\includegraphics[width=1.0\linewidth]{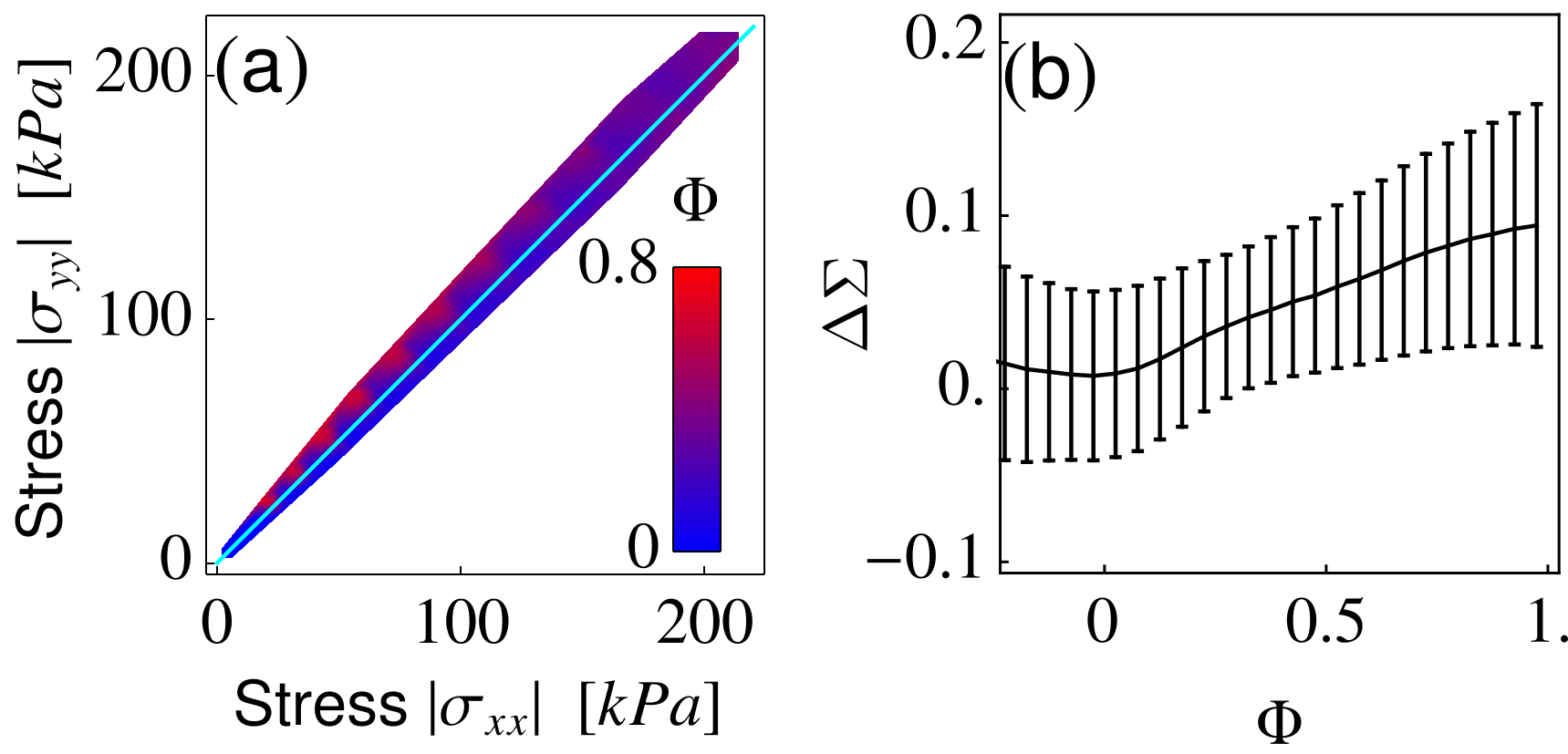}
\caption{\label{fig:stre-A} \textbf{Stress anisotropy in growing colonies.} \red{Relation between the stresses and the alignment parameter in colonies unconfined in the $y$-direction.} (a) Two-dimensional map of the anisotropy parameter $\Phi$ as a function of the normal stresses $\sigma_{xx}$ and $\sigma_{yy}$ \red{in channels of increasing width, in the range $L_{x}=50-150\,\mu$m with $\Delta L_{x}=10\,\mu$m.}. The \red{cyan} line corresponds to the bisectrix $|\sigma_{xx}|=|\sigma_{yy}|$. \red{The maxima of $\Phi$ correspond to the maxima of $\Delta\Sigma$ for a given $L_{x}$ value and their periodicity results solely from the combination multiple $L_{x}$ values, each associated with a maximum in $\Phi$ and $\Delta\Sigma$.} (b) Stress anisotropy parameter, Eq.~\eqref{eq:stress_anisotropy}, versus the alignment parameter. The error bars show the standard deviations of data samples about the average values.}
\end{figure}

In order to quantify such a correlation, we construct a two-dimensional map of the alignment parameter $\Phi$ as a function of $\sigma_{xx}$ and $\sigma_{yy}$ for the case of unconfined colonies (Fig.~\ref{fig:stre-A}a). This is achieved by varying the width of the channel in the interval $50\,\mu{\rm m}\le L_{x} \le 150\,\mu{\rm m}$ in steps of $10\ \mu$m and performing $100$ simulations for each $L_{x}$ value, with the goal of generating a large amount of $(\sigma_{xx},\sigma_{yy},\Phi)$ combinations. As it is evident from Fig.~\ref{fig:stre-A}a, the full data set only occupies a narrow wedge-shaped region of the $(\sigma_{xx},\sigma_{yy})-$plane, straddling the $|\sigma_{xx}|=|\sigma_{yy}|$ line, as a consequence of the stress redistribution resulting from the rotation of the cells. \red{Such a region is slightly asymmetric toward the $|\sigma_{yy}|>|\sigma_{xx}|$ half-plane because of the elongated morphology of the colony and the continual release of $\sigma_{xx}$ by the absorbing boundaries. This corresponds to a region of increased horizontal alignment, marked in red in Fig.~\ref{fig:stre-A}a, which is more likely to occur in the portions of the region characterized by large stress anisotropy.} The latter property is even more evident in Fig.~\ref{fig:stre-A}b, where the alignment parameter $\Phi$ is plotted directly against the stress anisotropy parameter, defined as:
\begin{equation}\label{eq:stress_anisotropy}
\Delta\Sigma = \frac{|\sigma_{yy}|-|\sigma_{xx}|}{|\sigma_{xx}|+|\sigma_{yy}|}\;,
\end{equation}
thus confirming our hypothesis that bacterial alignment is ultimately related with the occurrence of a confinement-induced anisotropy in the normal stresses.

\red{Finally, topological defects are observed in the early stages of the system for both discrete and continuous models. However their presence is transient and does not affect the long time behavior of the colony.}

\subsection{\label{sec:continuum}Continuum description}

\begin{figure}[t]
\centering
\includegraphics[width=1.0\linewidth]{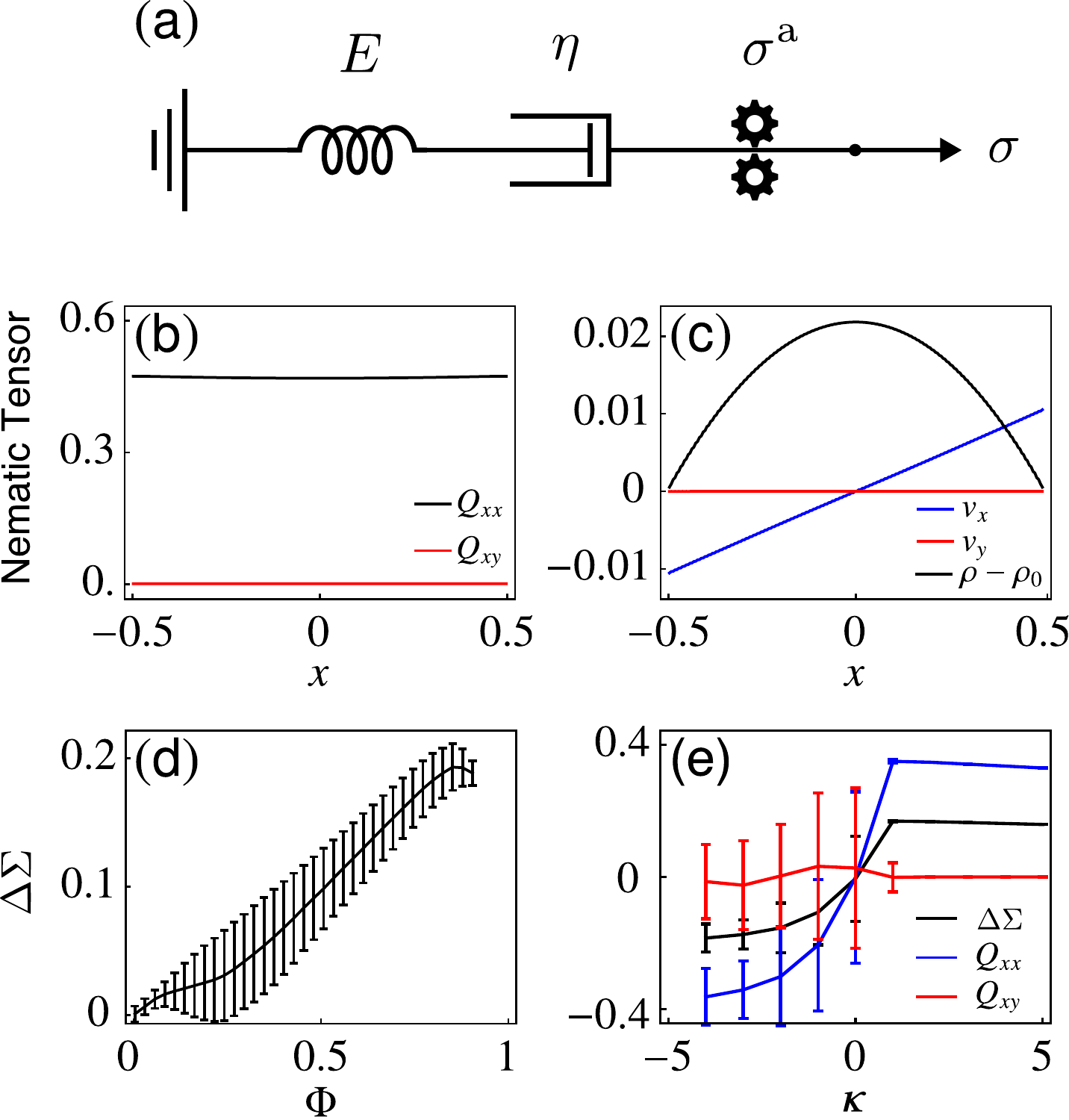}
\caption{\label{fig:Cont} \red{\textbf{Continuum model.}} (a) Schematic representation of a one-dimensional active Maxwell material. Activity is incorporated into the constitutive equation in the form of a generator supplying a constant stress $\sigma^{\rm a}$. (b) Steady state configurations of nematic tensor $\bm{Q}$ for $\kappa>0$. After an initial transient the colony relaxes toward a state characterized by perfect longitudinal alignment. (c) Steady state configuration of the density (black), $v_x$ (blue) and $v_y$ (red) in a colony with $\kappa>0$.  (d) Stress anisotropy parameter $\Delta\Sigma$, Eq.~\eqref{eq:stress_anisotropy}, versus the alignment parameter $\Phi$ for a colony with $\kappa>0$, compare with Fig.~\ref{fig:stre-A}b. (e) Steady state configurations of \red{$Q_{xx}$ (or $\Phi/2$), $Q_{xy}$}, and the stress anisotropy versus $\kappa$. As $\kappa$ changes in sign from positive to negative, the colony switches from longitudinal to transverse orientation. \red{The results are obtained from a numerical integration of Eqs.~\eqref{eq:hydro} on a rectangular domain with adsorbing boundaries on the $x-$direction and periodic boundaries on the $y-$direction.}}
\end{figure}
As a final test of the alignment mechanism proposed in the previous sections, we introduce a continuum model where most of the emergent properties found in our hard-rod model can be directly implemented and controlled. For this purpose we described monolayers of growing bacteria as two-dimensional viscoelastic active nematic gels, \red{characterized by three continuum fields: the cell density $\rho$, the local mechanical stress $\boldsymbol{\sigma}$, and the nematic tensor $\textbf{Q}$ indicating the local cell orientation. More details of the continuum model can be found in Sec. \ref{sec:contmodel}}.

Figs.~\ref{fig:Cont}b--d summarize the main outcome of the model when $\kappa>0$. After an initial transient, our continuum colony relaxes on a steady state characterized by perfect longitudinal alignment throughout the channel (Fig.~\ref{fig:Cont}b, where $Q_{xy}= S/2 \sin 2\theta = 0$ and $Q_{xx} = S/2 \cos 2\theta > 0$) and a stationary flow in the $x-$direction (Fig.~\ref{fig:Cont}c). By contrast, the density $\rho-\rho_{0}$ are maximal at $x=0$ and vanish at the outlets, as a consequence of the accumulation of cells in the center of the channel. A comparison between Fig.~\ref{fig:Cont}d and Fig.~\ref{fig:stre-A}b, in particular, corroborates our hypothesis that the faster build-up of transverse stress, originating from the inhibition of the flow along the $y-$direction, drives the longitudinal alignment of the nematic director. In Fig.~\ref{fig:Cont}d different combinations of $(\sigma_{xx},\sigma_{yy},\Phi)$ are sampled during the relaxation dynamics of the colony and the error bars indicate the standard deviation. 

Fig.~\ref{fig:Cont}e illustrates the effect of the coupling parameter $\kappa$ on the performance of global alignment. For each $\kappa$ value we run 50 simulations with different random initial configuration of the $\bm{Q}$ field to generate statistics. For $\kappa>0$, the colony rapidly evolves toward a longitudinally aligned configuration. The latter, in turn, is insensitive to the specific $\kappa$ value as exemplified by the plateau in $\Phi$ shown in Fig.~\ref{fig:Cont}e. By contrast, for $\kappa<0$, the colony exhibits global {\em vertical} alignment (i.e. $\Phi \approx -1$), but the performance of the alignment mechanism is reduced compared to $\kappa>0$ case, as demonstrated by the large error bars. We expect this behavior to originate from the incompatibility between vertical alignment and the release of active stress into the bacterial flow. While in the case of horizontally aligned cells the active stress $\bm{\sigma}^{\rm a} \sim \alpha \bm{\hat{x}}\bm{\hat{x}}$ originating from bacterial growth can be instantaneously released in the $x-$direction by sourcing a longitudinal expansion flow sinking at the absorbing boundaries, the active stress $\bm{\sigma}^{\rm a} \sim \alpha \bm{\hat{y}}\bm{\hat{y}}$ generated by vertically aligned cells must first be redistributed to the other stress component before being released through the flow.  

\begin{figure}[t!]
\centering
\includegraphics[width=1.0\linewidth]{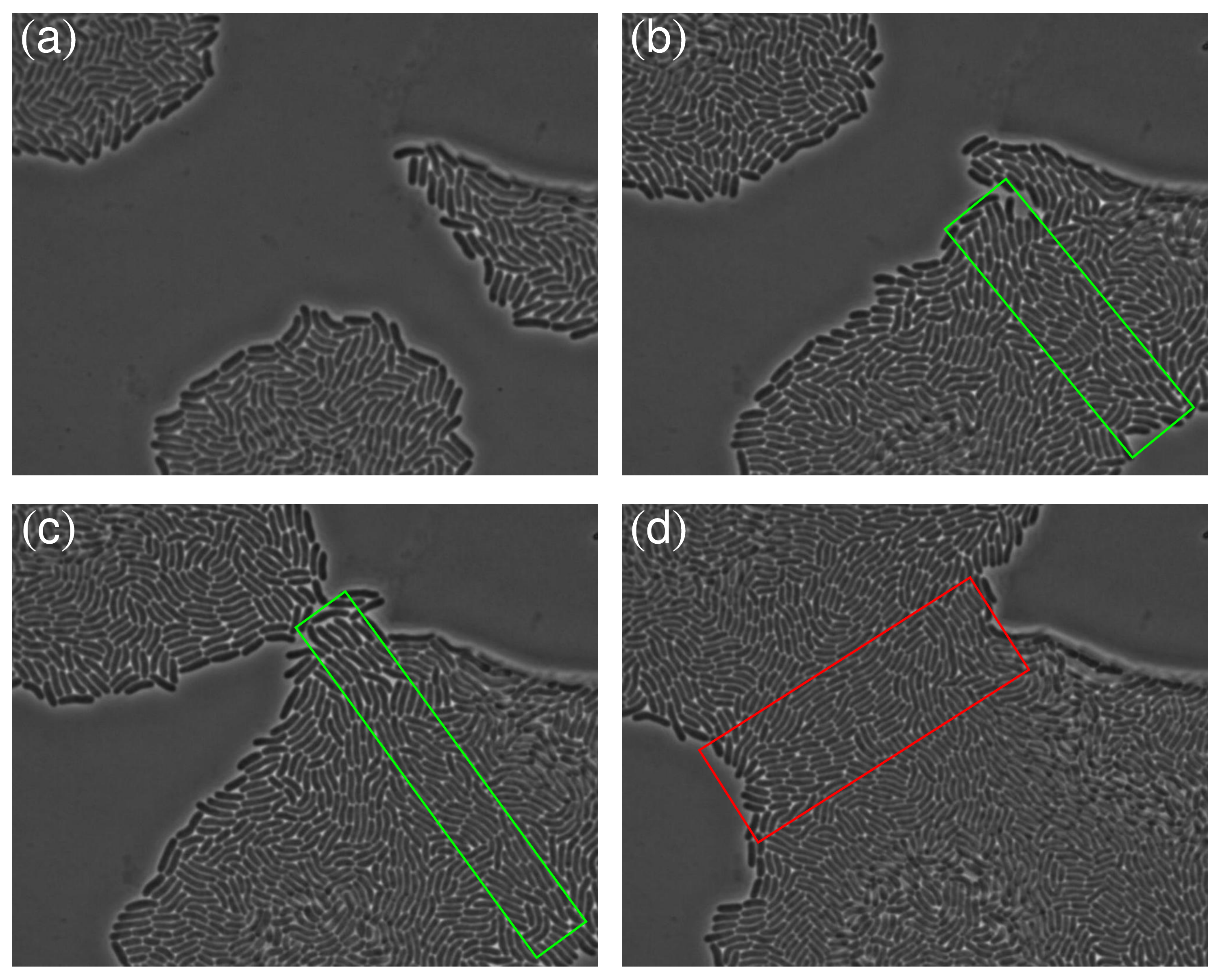}
\caption{\label{fig:co_merge} \textbf{Bacteria alignment in merging colonies.} \red{Experimental} snapshots of the merger of three {\em V. Cholerae} colonies at different times. In the merging regions, marked with colored boxes, cells collectively align with the common tangent at the colonies' front. Courtesy of Anupam Sengupta.} 
\end{figure}

\section{\label{sec:conclusions}Conclusions}

In this article we have investigated the interplay among orientational order, geometrical confinement and growth in multicellular systems of sessile bacteria. Using molecular dynamics and continuous modeling, we have demonstrated that geometrical anisotropies in the confining environment give rise to imbalance in the normal stresses, which, in turn, drives a collective reorientation of the cells. Cell reorientation not only eases the mechanical anisotropy by redistributing the existing stress among the two normal components, but also redirects the growth-induced flow in such a way to facilitate the escape of cells and reduce further stress build-up. The combination of these mechanisms results in a three-way regulatory loop that, although having a purely mechanical origin, allows bacterial colonies under confinement to control internal stresses and navigate through their environment in such a way to maximize their fitness. As an example of this behavior, here we discussed the case of bacterial colonies in a channel, where the fast build-up of transverse stress drives the cells to reorient along the longitudinal direction and proliferate throughout the channel. A chief prediction of our analysis is the relation between the stress anisotropy $\Delta \Sigma$ and alignment parameter $\Phi$ (see Figs.~\ref{fig:stre-A}b and \ref{fig:Cont}d), which could be, in principle, experimentally tested upon confining bacterial colonies with embedded stress sensors in a micro-channel.

Aside from the microfluidic setting, our work sheds light on special circumstances where confinement is induced by the colony itself, even in the absence of physical barriers. An interesting illustration of such a self-confinement effect, is provided by the merger of bacterial communities as they independently colonize the surrounding environment (Fig.~\ref{fig:co_merge}). In the merging region (enclosed by colored boxes), cells collectively align tangentially with respect to the colonies' front. This observation can be interpreted on the basis of our results by noticing that the cells at the interface between the merging colonies are confined by their respective bulks, thus subject to a normal stress $\sigma_{NN}=\bm{N}\cdot\bm{\sigma}\cdot\bm{N}$ that eventually exceeds the tangential stress $\sigma_{TT}=\bm{T}\cdot\bm{\sigma}\cdot\bm{T}$, where $\bm{N}$ and $\bm{T}$ are, respectively, the tangent and normal directions at the colony's front. Like colonies growing within an open channel, these interfacial cells are then reoriented in the tangential direction. This reorganization favors the merging of the bacterial communities and allow them to cooperatively colonize the surrounding space, rather than compete for the available resources. 

Our work additionally suggests a possible strategy to control the expansion trajectory of growing bacterial colonies with engineered anisotropic stresses. This could be achieved, for instance, by designing apposite confinement geometries or by introducing cell sinks at specific locations, in such a way to adjust the direction of minimal stress. These sinks could be, in principle, opened and closed on demand in order to achieve a fully controllable collective bacterial flow. 

Finally, our analysis progresses the current knowledge about hydrodynamic stability of active fluids, by highlighting the role of confinement and viscoelasticity as stabilizing features. It has been extensively shown, both theoretically and experimentally, that confined active fluids depart from the lowest free-energy state once the system size $L$ exceeds the characteristic length scale $\ell_{\rm a}$ at which active and passive forces and torques balance \cite{Simha:2002,Voituriez:2005,Marenduzzo:2007,Edwards:2009,Wioland:2016,Duclos:2018}. For $L \gg \ell_{\rm a}$, the system becomes chaotic \cite{Thampi:2013,Giomi:2015,Doostmohammadi:2017,Opathalage:2019,Lemma:2019}, often after having experienced an intermediate regime characterized by the appearance of coherent structures and periodic flow patterns \cite{Giomi:2011,Giomi:2012,Sumino:2012,Doostmohammadi:2016}. Here we showed that the combination of stress anisotropy and viscoelasticity enhances the stability of the lowest energy state, even in the presence of strong active flows, thus suggesting a new angle to explore the role of confinement in active matter.

\section{\label{sec:methods}Methods}
\subsection{\label{sec:hdmodel}Hard-rod model}

Cells are modeled as spherocylinders of a fixed diameter $d_{0}$ constrained to lie in the $xy-$plane. Each cell is characterized by three time-dependent degrees of freedom, namely the position $\bm{r}_{i}$, the orientation $\bm{p}_{i}=\cos\theta_{i}\,\bm{\hat{x}}+\sin\theta_{i}\,\bm{\hat{y}}$, with $\theta_{i}$ the angle with respect to the $x-$axis of a Cartesian frame, and the length $l_{i}$ (excluding the caps on both ends), \red{whose dynamics is governed by the following set of overdamped ordinary differential equations:
\begin{subequations}\label{eq:mo_drdq}
\begin{align}
\frac{{\rm d}\bm{r}_{i}}{{\rm d}t} &= \frac{1}{\zeta l_i}\,\left( \sum_{j=1}^{\mathcal{N}_{i}}\bm{F}^{c}_{ij} + \sum_{\alpha= \pm 1}\bm{F}^{w}_{i\alpha} + \boldsymbol{\eta}_{i}\right)\;,\\
\frac{{\rm d}\bm{p}_{i}}{{\rm d}t} &= \frac{12}{\zeta l_i^{3}}\,\bm{M}_{i}\times \bm{p}_{i}\;.
%\bm{M}_{i}&=\sum_{j=1}^{N_{i}^{c}}(\bm{r}_{ij}\times\bm{F}^{c}_{ij}) + \sum_{\alpha=\pm 1}(\bm{r}_{i\alpha}\times\bm{F}^{w}_{i\alpha})\;.   
\end{align}
\end{subequations}
The quantity $\zeta$ represents the effective drag per unit length originating from the cell-substrate adhesive interactions (see e.g. Ref. \cite{Duvernoy:2018}). Adhesive molecules expressed on the bacterial surface dynamically bind and unbind to the substrate and can be considered uniformly distributed along the cellular wall on average. The resulting drag force is, consequently, proportional to the cell area and is independent of the cell orientation. The first term on the right-hand side of Eq. (\ref{eq:mo_drdq}a) represents the steric force due to the contact between the $i-$th and $j-$th cell, with $\mathcal{N}_{i}$ the total number of cells in contact with the $i-$th. The points of contact have positions $\bm{r}_{ij}$ with respect to the center of mass of the $i-$th cell and apply a Hertzian force 
\begin{equation}
\bm{F}_{ij}^{c}=Yd_{0}^{1/2}h_{ij}^{3/2}\bm{N}_{ij}\;, 
\end{equation}
where $Y$ is proportional to the Young’s modulus of the cell, $h_{ij}$ is the overlap distance between the two cells, and $\bm{N}_{ij}$ their common normal. The second term on the right-hand side of Eq. (\ref{eq:mo_drdq}a), on the other hand, represents the force associated with the interaction between the cell caps, at positions $\bm{r}_{i\alpha}=\alpha l_{i}\bm{p}_{i}/2$ with respect to the center of mass, and the confining walls and is calculated analogously to $\bm{F}_{ij}^{c}$. The vector $\bm{\eta}_{i}$ corresponds to a random force, whose components are sampled from the uniform distribution in the range $-10^{-6}\,{\rm N}\le \eta_{i\beta}\le 10^{-6}\,{\rm N}$, with $\beta=x,y$. This force embodies the inevitable randomization of the bacterial dynamics resulting from small irregularities in the cell walls and the substrate, preventing the dividing cells from forming long one-dimensional chains that are never observed in experiments. Analogously, Eq. (\ref{eq:mo_drdq}b) describes the rotation of the cell axis in response to the torque: 
\begin{equation}
\bm{M}_{i}=\sum_{j=1}^{\mathcal{N}_{i}}(\bm{r}_{ij}\times\bm{F}^{c}_{ij}) + \sum_{\alpha=\pm 1}(\bm{r}_{i\alpha}\times\bm{F}^{w}_{i\alpha})\;.   	
\end{equation}	
Finally, the length $l_i$ increases linearly in time with rate $g_{i}$ and, after having reached the division length $l_{d}$, the cell divides into two identical daughter cells. In order to avoid synchronous divisions, the growth rate is randomly sampled from the uniform distribution in the range $g/2 \le g_{i} \le 3g/2$, with $g$ the average growth rate. Immediately after duplication, the daughter cells have the same orientation as the mother cell, but independent growth rates. Eqs. \eqref{eq:mo_drdq} have been numerically integrated using the following set of parameter values, corresponding to the typical values of {\em E-coli} micro-colonies under physiological conditions (see e.g. Ref. \cite{Farrell:2013}): $d_{0}=1\,\mu$m, $l_{d}=2\,\mu$m, $g=2\,\mu$m/h, $Y=\,4$~MPa, and $\zeta=200$~Pa~h. The integration is performed via the Euler scheme with a time step $\Delta t=0.5\times 10^{-6}$~h.}

The proliferating colony is confined in a rectangular $L_{x} \times L_{y}$ box, whose horizontal \red{(top and bottom)} boundaries are either consisting of impenetrable rigid walls, periodic, \red{or placed at an infinite distance from the colony}. The vertical \red{(left and right)} boundaries, on the other hand, are always set as absorbing, in such a way to emulate the outlets of a microfluidic channel \cite{Volfson:2008, Cho:2007, Sheats:2017, Boyer:2011, Fuentes:2013}. Cells crossing the absorbing boundaries are then removed from the system. For the sake of brevity, in the remainder of the paper, we refer to $L_{x}$ as the ``width'' and to $L_{y}$ as the ``height'' of the channel. All simulations start with a cell at the origin, with random orientation. We change only the boundary condition and the channel geometry, and fix all the other parameters.

\subsection{\label{sec:contmodel}Continuum  model}
\red{We describe} monolayers of growing bacteria as two-dimensional viscoelastic active nematic gels, characterized by a Maxwell-like stress relaxation (Fig.~\ref{fig:Cont}a), namely:
\begin{equation}
  \label{eq:maxwell}
2 \eta \bm{u} = (1+\tau D_{t})(\bm{\sigma}-\bm{\sigma}^{\rm a})\;,
\end{equation}
where $u_{ij}=(\partial_{i}v_{j}+\partial_{j}v_{i})/2$ is the strain-rate tensor and the constants $\eta$ and $\tau$ represent, respectively, the shear viscosity and the viscoelastic crossover time: i.e. $\tau=\eta/E$, with $E$ the Young modulus of the colony. The operator $D_{t}$ is the corotational material derivative: i.e. $D_{t}\bm{\sigma}=\partial_{t}\bm{\sigma}+\bm{v}\cdot\nabla\bm{\sigma}+\bm{\omega}\cdot\bm{\sigma}-\bm{\sigma}\cdot\bm{\omega}$, with $\omega_{ij}=(\partial_{i}v_{j}-\partial_{j}v_{i})/2$ the vorticity tensor. The tensor $\bm{\sigma}^{\rm a}=\alpha\bm{Q}$ represents the extensile {\em active} stress resulting from the growth of the cells \cite{Pedley:1992,Simha:2002}, whereas $\bm{Q}=\langle \bm{p}\bm{p}-\mathbb{1}/2\rangle=S(\bm{n}\bm{n}-\mathbb{1}/2)$ is the nematic tensor, with $\mathbb{1}$ the two-dimensional identity tensor, $\bm{n}$ the nematic director, representing the average orientation of the cells and $S=\langle 2|\bm{n}\cdot\bm{p}|-1\rangle$ the local order parameter. At long times i.e. $t \gg \tau$, Eq.~\eqref{eq:maxwell} yields the usual constitutive equation of active fluids: i.e. $\bm{\sigma}=2\eta\bm{u}+\bm{\sigma}^{\rm a}$; while at short times, Eq.~\eqref{eq:maxwell} describes the characteristic response of a solid material: i.e. $\partial_{t}\bm{\sigma} \sim \bm{u}$. 

The hydrodynamic equations for the cell density $\rho$, momentum density $\rho\bm{v}$ and nematic tensor $\bm{Q}$, are given by:
\begin{subequations}\label{eq:hydro}
\begin{gather}
\partial_{t}\rho+\nabla\cdot(\rho \bm{v}) = k_{g}\rho\;,\\[5pt]
\partial_{t}(\rho \bm{v})+\nabla\cdot(\rho \bm{v}\bm{v}) = \nabla\cdot\bm{\tilde{\sigma}}-\nabla P -\xi \rho \bm{v}\;,\\[5pt]
D_{t}\bm{Q} = \gamma^{-1}\bm{\tilde{H}}\;,
\end{gather}
\end{subequations}	
where the tilde indicates the deviatoric (i.e. traceless) component: e.g. $\bm{\tilde{\sigma}}=\bm{\sigma}-\tr(\bm{\sigma})\mathbb{1}/2$.

Eq.~(\ref{eq:hydro}a) accounts for the exponential growth in the number of bacteria resulting from cell division, with $k_{g}$ the growth rate: i.e. $N=\int {\rm d}A\,\rho/m=N_{0}\exp k_{g}t$, with $m$ the cell mass. On the other hand, Eq.~(\ref{eq:hydro}b) describes the redistribution of momentum among the stresses, with $P$ the pressure, as well as its dissipation via the drag force $-\xi\rho\bm{v}$, \red{with $\xi$ a drag coefficient}, resulting from the interaction with the substrate. Finally, Eq.~(\ref{eq:hydro}c) governs the evolution of the bacterial orientational degrees of freedom, with $\gamma$ the orientational viscosity and $\bm{H}=-\delta F/\delta \bm{Q}$ the molecular tensor describing the relaxation of the free energy $F=\int {\rm d}A\,f$. The latter must account for both the entropic contribution, resulting from a departure from the uniformly aligned configuration, as well as the elastic contribution associated with the colony solid-like behavior and can be expressed in the classic de Gennes' form \cite{DeGennes:1975}: 
\begin{equation}
f = \frac{1}{2}K|\nabla\bm{Q}|^{2}+\frac{1}{2}A\tr\bm{Q^{2}}+\frac{1}{4}C(\tr\bm{Q}^{2})^{2}-\Gamma\tr(\bm{Q}\cdot\bm{\epsilon})\;,	
\end{equation}
with $K$ the orientational stiffness of the nematic phase, in one elastic constant approximation, and $A$, $C$ being mean-field coefficients controlling the isotropic-nematic phase transition. \red{As bacterial colonies are generally not subject to thermal fluctuations and nematic order arises as a consequence of crowding and activity (see e.g. Ref. \cite{DellArciprete:2018}), one can assume the order parameter $S=\sqrt{2\tr\bm{Q}^{2}}$ to be non-zero even in two dimensions. The constant} $\Gamma$ \red{expresses} the strength of the coupling between the nematic tensor and the strain tensor $\bm{\epsilon}$. Consistently with Eq.~\eqref{eq:maxwell}, we assume this to be related to the stress tensor via Hooke's law: i.e. $\bm{\epsilon}=(1+\nu)/E\,\bm{\sigma}-\nu/E\,\tr(\bm{\sigma})\mathbb{1}$, with $\nu$ the Poisson ratio. Thus:
\begin{equation}\label{eq:molecular_tensor}
\tilde{\bm{H}}=K\nabla^{2}\bm{Q}+\left(A+\frac{1}{2}CS^{2}\right)\bm{Q}-\kappa \bm{\tilde{\sigma}}\;,
\end{equation}
where $\kappa=\Gamma(1+\nu)/E$. All the coefficients appearing in Eqs.~(\ref{eq:hydro}b,c) depend, in principle, on the cell density $\rho$. In particular, here we assume the following simple linear dependence:
\[
S_{0}=\sqrt{\frac{-2A}{C}} = \sqrt{1-\frac{\rho_{c}}{\rho}}\;,\qquad
\alpha = \alpha_{0}\left(1-\frac{\rho}{\rho_{0}}\right)\;.
\]
Here $\rho_{c}$ is the critical density associated with the isotropic-nematic phase transition, whereas $\rho_{0}>\rho_{c}$, represents the density at which the cells a closely packed and able to transfer stresses on their surroundings. In addition, we assume the equation of state $P=P_{0}(\rho/\rho_{0}-1)$. 

The continuum model outlined here is closely related with the active gel model introduced by Kruse {\em et al}. to explain the formation of defective patterns in polar actomyosin gels \cite{Kruse:2004}, but further accounts for the remodeling caused elastic deformations of the bacterial domains at short time scales. \red{Experimental estimates of the viscoelastic crossover time $\tau$ are available in the literature on biofilms (see e.g. Refs \cite{Charlton:2019,Gloag:2020} for recent publications), but are less common for microcolonies, as the small size and the monolayer structure of these bacterial communities render standard rheological techniques ineffective. Yet, from Refs. \cite{Volfson:2008,DellArciprete:2018} one can infer $\tau$ to be of order $100$ min for {\em E. coli} under physiological conditions in confined and freely expanding colonies respectively, thus comparable with our molecular dynamics simulations (see Fig. \ref{fig:stre-cent}). Furthermore, Eqs. \eqref{eq:maxwell} and \eqref{eq:hydro}, do not account for rotational viscoelasticity, such as that encountered in certain type of polymeric liquid crystals, such as dense suspensions of actin filaments, and originating from the entanglement of these semiflexible polymers in three dimensions \cite{Julicher:2007}.}

Eqs.~\eqref{eq:hydro} are numerically integrated using finite differences on a rectangular domain endowed with stress-free absorbing boundaries on the $x-$direction and periodic boundaries on the $y-$direction. The system is initiated with $\rho=\rho_{0}$, $\bm{v}=\bm{0}$ and a random configuration of the $\bm{Q}$ field and evolved until a steady state is found. More details about the numerical integration of Eqs.~\eqref{eq:hydro}, including all parameter values, can be found in the Supplementary Information.

% --------------------------------------------------------------------------

% --------------------------------------------------------------------------

% ---------------------------------------------------------------------------
\acknowledgments

\noindent
\textbf{Funding:} We are indebted with Anupam Sengupta for several illuminating discussions and for sharing with us the images displayed in Fig.~\ref{fig:co_merge}. This work is partially supported by The Netherlands Organization for Scientific Research (NWO/OCW) as part of the Frontiers of Nanoscience program and the Vidi Scheme (ZY, DJGP and LG). \red{\textbf{Author contributions:} Z.Y., D.J.G.P., and L.G. developed the models, performed the simulations, analyzed the data, and wrote the manuscript.} \red{\textbf{Competing interests:} All authors declare that they have no competing interests.} \red{\textbf{Data and materials availability:} All data needed to evaluate the conclusions in the paper are present in the paper and/or the Supplementary Materials.}


\begin{thebibliography}{99}

\bibitem{Friedl:2004}
P. Friedl, Y. Hegerfeldt, M. Tusch,
Collective cell migration in morphogenesis and cancer.
\href{https://doi.org/10.1387/ijdb.041821pf}{Int J Dev Biol. {\bf 48}, 441--449 (2004)}.

\bibitem{Haeger:2015}
A. Haeger, K. Wolf, M. M. Zegers, P. Friedl,
Collective cell migration: guidance principles and hierarchies.
\href{https://doi.org/10.1016/j.tcb.2015.06.003}{Trends Cell Biol. {\bf 25}, 556--566 (2015)}.

\bibitem{BenJacob:1998}
E. Ben-Jacob, I. Cohen, D. L. Gutnick,
Cooperative organization of bacterial colonies: from genotype to morphotype.
\href{https://doi.org/10.1146/annurev.micro.52.1.779}{Annu. Rev. Microbiol. {\bf 52}, 779--806 (1998)}.

\bibitem{BenJacob:2000}
E. Ben-Jacob, I. Cohen, H. Levine,
Cooperative self-organization of microorganisms.
\href{https://doi.org/10.1080/000187300405228}{Adv. Phys. {\bf 49}, 395--554 (2000)}.

\bibitem{Allen:2018}
R. J. Allen, B. Waclaw,
Bacterial growth: a statistical physicist's guide.
\href{https://doi.org/10.1088/1361-6633/aae546}{Rep. Prog. Phys. {\bf 82}, 016601 (2018)}.

\bibitem{Volfson:2008}	
D. Volfson, S. Cookson, J. Hasty, L. S. Tsimring,
Biomechanical ordering of dense cell populations.
\href{http://dx.doi.org/doi:10.1073/pnas.0706805105}{Proc. Natl. Acad. Sci. U. S. A. {\bf 105}, 15346--15351 (2008)}.

\bibitem{Boyer:2011}	
D. Boyer, W. Mather, O. Mondrag\'{o}n-Palomino, S. Orozco-Fuentes, T. Danino, J. Hasty, L. S. Tsimring,
{\em Buckling instability in ordered bacterial colonies}.
\href{http://dx.doi.org/doi:10.1088/1478-3975/8/2/026008}{Phys. Biol. {\bf 8}, 026008 (2011)}.

\bibitem{Karamched:2019}
B.R. Karamched, W. Ott, I. Timofeyev, R.N. Alnahhas, M.R. Bennett, K. Josi{\'c},
Moran model of spatial alignment in microbial colonies.
\href{https://doi.org/10.1016/j.physd.2019.02.001}{Physica D {\bf 395}, 1--6 (2019)}.

\bibitem{Cho:2007}	
H. Cho, H. J\"{o}nsson, K. Campbell, P. Melke, J. W. Williams, B. Jedynak, A. M. Stevens, A. Groisman, A. Levchenko,
Self-organization in high-density bacterial colonies: efficient crowd control.
\href{http://dx.doi.org/doi:10.1371/journal.pbio.0050302}{PLoS Biol. {\bf 5}, e302 (2007)}.

\bibitem{Fuentes:2013}	
S. Orozco-Fuentes, D. Boyer,
Order, intermittency, and pressure fluctuations in a system of proliferating rods.
\href{http://dx.doi.org/doi:10.1103/PhysRevE.88.012715}{Phys. Rev. E {\bf 88}, 012715 (2013)}.

\bibitem{Sheats:2017}
J. Sheats, B. Sclavi, M. C. Lagomarsino, P. Cicuta, K. D. Dorfman,
Role of growth rate on the orientational alignment of \textit{Escherichia Coli} in a slit.
\href{http://rsos.royalsocietypublishing.org/content/4/6/170463}{R. Soc. Open Sci. {\bf 4}, 170463 (2017)}.

\bibitem{Farrell:2013}	
F. D. C. Farrell, O. Hallatschek, D. Marenduzzo, B. Waclaw,
Mechanically driven growth of quasi-two-dimensional microbial colonies.
\href{http://dx.doi.org/doi:10.1103/PhysRevLett.111.168101}{Phys. Rev. Lett. {\bf 111}, 168101 (2013)}.

\bibitem{Grant:2014}	
M. A. A. Grant, B. Waclaw, R. J. Allen, and P. Cicuta,
The role of mechanical forces in the planar-to-bulk transition in growing \textit{Escherichia Coli} microcolonies.
\href{http://dx.doi.org/10.1098/rsif.2014.0400}{J. R. Soc. Interface {\bf 11}, 20140400 (2014)}.

\bibitem{You:2018}	
Z. You, D. J. G. Pearce, A. Sengupta, L. Giomi,
Geometry and mechanics of microdomains in growing bacterial colonies.
\href{http://dx.doi.org/doi:10.1103/PhysRevX.8.031065}{Phys. Rev. X {\bf 8}, 031065 (2018)}.

\bibitem{Duvernoy:2018}
M.-C. Duvernoy, T. Mora, M. Ardr\`e, V. Croquette, D. Bensimon, C. Quilliet, J.-M. Ghigo, M. Balland, C. Beloin, S. Lecuyer, N. Desprat, 
Asymmetric adhesion of rod-shaped bacteria controls microcolony morphogenesis.
\href{https://doi.org/10.1038/s41467-018-03446-y}{Nat. Commun. {\bf 9}, 1120 (2018)}.

\bibitem{Winkle:2017}
J. J. Winkle, O. A. Igoshin, M. R. Bennett, K. Josi{\'c}, W. Ott,
Modeling mechanical interactions in growing populations of rod-shaped bacteria.
\href{https://doi.org/10.1088/1478-3975/aa7bae}{Phys. Bio. {\bf 14}(5), 055001 (2017)}.

\bibitem{Alnahhas:2019}
R. N. Alnahhas, J. J. Winkle, A. J. Hirning, B. Karamched, W. Ott, K. Josi{\'c}, M. R. Bennett,
Spatiotemporal dynamics of synthetic microbial consortia in microfluidic devices.
\href{https://doi.org/10.1021/acssynbio.9b00146}{ACS Syn. Bio. {\bf 8}(9), 2051--2058 (2019)}.

\bibitem{DeGennes:1993}
P. G. de Gennes, J. Prost,
The Physics of Liquid Crystals, 2nd ed.
(Oxford University Press, Oxford 1993).

%\bibitem{Duclos:2017}
%G. Duclos, C. Erlenk{\"a}mper, J.-F. Joanny, P. Silberzan,
%Topological defects in confined populations of spindle-shaped cells.
%\href{https://doi.org/10.1038/nphys3876}{Nat. Phys. {\bf 13}, 58--62 (2017)}.

\bibitem{Duclos:2014}
G. Duclos, S. Garcia, H. G. Yevick, P. Silberzan,
Perfect nematic order in confined monolayers of spindle-shaped cells.
\href{https://doi.org/10.1039/c3sm52323c}{Soft Matter {\bf 10}, 2346--2353 (2014)}.

\bibitem{DellArciprete:2018}
D. Dell’Arciprete, M. L. Blow, A. T. Brown, F. D. C. Farrell, J. S. Lintuvuori, A. F. McVey, D. Marenduzzo, W. C. K. Poon,
A growing bacterial colony in two dimensions as an active nematic.,
\href{https://doi.org/10.1038/s41467-018-06370-3}{Nat. Commun. {\bf 9}(1), 4190 (2018)}.

%\bibitem{Francescangeli:2010}
%O. Francescangeli, E. T. Samulski,  
%Insights into the cybotactic nematic phase of bent-core molecules.
%\href{https://doi.org/10.1039/C003310C}{Soft Matter {\bf 6}, 2413--2420 (2010)}.

\bibitem{DeVries:1970}
A. De Vries,
Evidence for the existence of more than one type of nematic phase,
\href{https://doi.org/10.1080/15421407008083484}{Mol. Cryst. Liq. Cryst. {\bf 10}, 31--35 (1970)}. 

\bibitem{Frenkel:1987}
D. Frenkel,
Onsager's spherocylinders revisited.
\href{https://doi.org/10.1021/j100303a008}{J. Phys. Chem. {\bf 91}, 4912--4916 (1987)}.

\bibitem{Sauer:2016}
M. M. Sauer, R. P. Jakob, J. Eras, S. Baday, D. Eriş, G. Navarra, S. Bern\`eche, B. Ernst, T. Maier, R. Glockshuber, 
Catch-bond mechanism of the bacterial adhesin FimH.
\href{https://doi.org/10.1038/ncomms10738}{Nat. Commun. {\bf 7}, 10738 (2016)}.

\bibitem{Pedley:1992}
T. J. Pedley, J. O. Kessler,
Hydrodynamic phenomena in suspensions of swimming microorganisms.
\href{http://dx.doi.org/10.1146/annurev.fl.24.010192.001525}{Ann. Rev. Fluid Mech. {\bf 24}, 313--358 (1992)}.

\bibitem{Simha:2002}
R. A. Simha, S. Ramaswamy,
Hydrodynamic fluctuations and instabilities in ordered suspensions of self-propelled particles.
\href{https://doi.org/10.1103/PhysRevLett.89.058101}{Phys. Rev. Lett. {\bf 89}, 058101 (2002)}.

\bibitem{DeGennes:1975}
P. G. de Gennes,
{\em Reflexions sur un type de polymeres nematiques},
C. R. Acad. Sci. {\bf B281}, 101 (1975).

\bibitem{Kruse:2004}
K. Kruse,  J.F. Joanny, F. J{\"u}licher, J. Prost, K. Sekimoto,
Asters, vortices, and rotating spirals in active gels of polar filaments.
\href{https://doi.org/10.1103/PhysRevLett.92.078101}{Phys. Rev. Lett. {\bf 92}, 078101 (2004)}.

\bibitem{Charlton:2019}
S. G. V. Charlton, M. A. White, S. Jana, L. E. Eland, P. G. Jayathilake, J. G. Burgess, J. Chen, A. Wipat, T. P. Curtis,
Regulating, measuring, and modeling the viscoelasticity of bacterial biofilms.
\href{https://doi.org/10.1128/JB.00101-19}{J. Bacteriol. {\bf 201}, 00101 (2019)}.

\bibitem{Gloag:2020}
E. S. Gloag, S. Fabbri, D. J.Wozniak, P. Stoodley,
Biofilm mechanics: Implications in infection and survival.
\href{https://doi.org/10.1016/j.bioflm.2019.100017}{Biofilm {\bf 2}, 100017 (2020)}.

\bibitem{Julicher:2007}
F. J\"ulicher, K. Kruse, J. Prost, J.-F. Joanny,
Active behavior of the cytoskeleton.
\href{https://doi.org/10.1016/j.physrep.2007.02.018}{Phys. Rep. {\bf 449}, 3 (2007)}. 

\bibitem{Voituriez:2005}
R. Voituriez, J.-F. Joanny, J. Prost,
Spontaneous flow transition in active polar gels.
\href{http://dx.doi.org/10.1209/epl/i2004-10501-2}{Europhys. Lett. {\bf 70}, 404--410 (2005)}.

\bibitem{Marenduzzo:2007}
D. Marenduzzo, E. Orlandini, M. E. Cates, J. M. Yeomans,
Steady-state hydrodynamic instabilities of active liquid crystals: hybrid lattice Boltzmann simulations.
\href{http://dx.doi.org/10.1103/PhysRevE.76.031921}{Phys. Rev. E {\bf 76}, 031921 (2007)}.

\bibitem{Edwards:2009}
S. A. Edwards, J. M. Yeomans,
Spontaneous flow states in active nematics: a unified picture.
\href{http://dx.doi.org/10.1209/0295-5075/85/18008}{Europhys. Lett. {\bf 85}, 18008 (2009)}.

\bibitem{Wioland:2016}
H. Wioland, E. Lushi, R. E. Goldstein,
Directed collective motion of bacteria under channel confinement.
\href{http://dx.doi.org/10.1088/1367-2630/18/7/075002}{New J. Phys. {\bf 18} 075002 (2016)}.

\bibitem{Duclos:2018}
G. Duclos, C. Blanch-Mercader, V. Yashunsky, G. Salbreux, J.-F. Joanny, J. Prost, P. Silberzan,
Spontaneous shear flow in confined cellular nematics.
\href{https://doi.org/10.1038/s41567-018-0099-7}{Nat. Phys. {\bf 14}, 728--732 (2018)}.

\bibitem{Thampi:2013}
S. P. Thampi, R. Golestanian, J. M. Yeomans,
Velocity correlations in an active nematic.
\href{https://doi.org/10.1103/PhysRevLett.111.118101}{Phys. Rev. Lett. {\bf 111}, 118101 (2013)}.

\bibitem{Giomi:2015}
L. Giomi,
Geometry and topology of turbulence in active nematics.
\href{http://dx.doi.org/10.1103/PhysRevX.5.031003}{Phys. Rev. X {\bf 5}, 031003 (2015)}.

\bibitem{Doostmohammadi:2017}
A. Doostmohammadi, T. N. Shendruk, K. Thijssen, J. M. Yeomans,
Onset of meso-scale turbulence in active nematics.
\href{http://dx.doi.org/10.1038/ncomms15326}{Nat. Commun. {\bf 8}, 15326 (2017)}.

\bibitem{Opathalage:2019}
A. Opathalage, M. M. Norton, M. P. N. Juniper, B. Langeslay, S. A. Aghvami, S. Fraden, Z. Dogic,
Self-organized dynamics and the transition to turbulence of confined active nematics.
\href{https://doi.org/10.1073/pnas.1816733116}{Proc. Nat. Acad. Sci. U. S. A. {\bf 116}, 4788--4797 (2019)}.

\bibitem{Lemma:2019}
L. M. Lemma, S. J. DeCamp, Z. You, L. Giomi, Z. Dogic,
Statistical properties of autonomous flows in 2D active nematics
\href{https://doi.org/10.1039/C8SM01877D}{Soft Matter {\bf 15}, 3264--3272 (2019)}.

\bibitem{Sumino:2012}	
Y. Sumino, K. H. Nagai, Y. Shitaka, D. Tanaka,	K. Yoshikawa, H. Chat\'e, K. Oiwa,
Large-scale vortex lattice emerging from collectively moving microtubules.
\href{http://dx.doi.org/10.1038/nature10874}{Nature {\bf 483}, 448–-452 (2012)}.

\bibitem{Giomi:2011}
L. Giomi, L. Mahadevan, B. Chakraborty, M. F. Hagan,
Excitable patterns in active nematics.
\href{http://prl.aps.org/abstract/PRL/v106/i21/e218101}{Phys. Rev. Lett. {\bf 106}, 218101 (2011)}.

\bibitem{Giomi:2012}
L. Giomi, L. Mahadevan, B. Chakraborty, M. F. Hagan,
Banding, excitability and chaos in active nematic suspensions.
\href{http://dx.doi.org/10.1088/0951-7715/25/8/2245}{Nonlinearity {\bf 25}, 2245--2269 (2012)}.

\bibitem{Doostmohammadi:2016}
A. Doostmohammadi, M. F. Adamer, S. P. Thampi, J. M. Yeomans,
Stabilization of active matter by flow-vortex lattices and defect ordering.
\href{http://dx.doi.org/10.1038/ncomms10557}{Nat. Commun. {\bf 7}, 10557 (2016)}.

%\bibitem{Mcdougald:2012}	
%D. McDougald, S. A. Rice, N. Barraud, P. D. Steinberg, S. Kjelleberg,
%Should we stay or should we go: mechanisms and ecological consequences for biofilm dispersal.
%\href{http://dx.doi.org/doi:10.1038/nrmicro2695}{Nat. Rev. Microbiol. {\bf 10}, 39--50 (2012)}.

%\bibitem{Doostmohammadi:2016}
%A. Doostmohammadi, S. P. Thampi, J. M. Yeomans,
%Defect-mediated morphologies in growing cell colonies.
%\href{https://doi.org/10.1103/PhysRevLett.117.048102}{Phys. Rev. Lett. {\bf 117}, 048102 (2016)}.

%\bibitem{Rosan:2000}	
%B. Rosan, R. J. Lamont,
%Dental plaque formation.
%\href{http://dx.doi.org/doi:10.1016/S1286-4579(00)01316-2}{Microb. Infect. {\bf 2}, 1599--1607 (2000)}.

%\bibitem{Kaplan:2010}	
%J. B. Kaplan,
%Biofilm dispersal: mechanisms, clinical implications, and potential therapeutic uses.
%\href{http://dx.doi.org/doi:10.1177/0022034509359403}{J. Dent. Res. {\bf 89}, 205--218 (2010)}.

%\bibitem{Costerton:1999}	
%J. W. Costerton, P. S. Stewart, E. P. Greenberg,
%Bacterial biofilms: a common cause of persistent infections.
%\href{http://dx.doi.org/doi:10.1126/science.284.5418.1318}{Science {\bf 284}, 1318--1322 (1999)}.

%\bibitem{Costerton:1995}	
%J. W. Costerton, Z. Lewandowski, D. E. Caldwell, D. R. Korber, H. M. Lappin-Scott,
%Microbial biofilms.
%\href{http://dx.doi.org/doi:10.1146/annurev.mi.49.100195.003431}{Ann. Rev. Microbiol. {\bf 49}, 711--745 (1995)}.

%\bibitem{Persat:2015}	
%A. Persat, C. D. Nadell, M. K. Kim, F. Ingremeau, A. Siryaporn, K. Drescher, N. S. Wingreen, B. L. Bassler, Z. Gitai, H. A. Stone,
%The mechanical world of bacteria.
%\href{http://dx.doi.org/doi:10.1016/j.cell.2015.05.005}{Cell {\bf 161}, 988--997 (2015)}.

%\bibitem{Hoffman:2011}	
%B. D. Hoffman, C. Grashoff, M. A. Schwartz,
%Dynamic molecular processes mediate cellular mechanotransduction.
%\href{http://dx.doi.org/doi:10.1038/nature10316}{Nature {\bf 475}, 316--323 (2011)}.

%\bibitem{Morris:2008}	
%D. M. Morris, G. J. Jensen,
%Toward a biomechanical understanding of whole bacterial cells.
%\href{http://dx.doi.org/doi:10.1146/annurev.biochem.77.061206.173846}{Ann. Rev. Biochem. {\bf 77}, 583--613 (2008)}.

%\bibitem{Su:2012}	
%P.-T. Su, C.-T. Liao, J.-R. Roan, S.-H. Wang, A. Chiou, W.-J. Syu,
%Bacterial colony from two-dimensional division to three-dimensional development.
%\href{http://dx.doi.org/doi:10.1371/journal.pone.0048098}{PLoS ONE {\bf 7}, e48098 (2012)}.

%\bibitem{Rudge:2013}	
%T. J. Rudge, F. Federici, P. J. Steiner, A. Kan, J. Haseloff,
%Cell polarity-driven instability generates self-organized, fractal patterning of cell layers.
%\href{http://dx.doi.org/doi:10.1021/sb400030p}{ACS Synth. Biol. {\bf 2}, 705--714 (2013)}.

%\bibitem{Farrell:2013}	
%F. D. C. Farrell, O. Hallatschek, D. Marenduzzo, B. Waclaw,
%Mechanically driven growth of quasi-two-dimensional microbial colonies.
%\href{http://dx.doi.org/doi:10.1103/PhysRevLett.111.168101}{Phys. Rev. Lett. {\bf 111}, 168101 (2013)}.

%\bibitem{Grant:2014}	
%M. A. A. Grant, B. Waclaw, R. J. Allen, and P. Cicuta,
%The role of mechanical forces in the planar-to-bulk transition in growing \textit{Escherichia Coli} microcolonies.
%\href{http://dx.doi.org/10.1098/rsif.2014.0400}{J. R. Soc. Interface {\bf 11}, 20140400 (2014)}.

%\bibitem{Sanchez:2012}
%T. Sanchez, D. N. Chen, S. J. DeCamp, M. Heymann, and Z. Dogic,
%Spontaneous motion in hierarchically assembled active matter.
%\href{http://dx.doi.org/10.1038/nature11591}{Nature {\bf 491}, 431--434 (2012)}.

%\bibitem{DeCamp:2015}
%S. J. DeCamp, G. S. Redner, A. Baskaran, M. F. Hagan, Z. Dogic,
%Orientational order of motile defects in active nematics.
%\href{https://doi.org/10.1038/nmat4387}{Nat. Mater. {\bf 14}, 1110--1115 (2015)}.
		
%\bibitem{Guillamat:2016}
%P. Guillamat, J. Ign\'{e}s-Mullol, F. Sagu\'{e}s,
%Control of active liquid crystals with a magnetic field.
%\href{http://dx.doi.org/10.1073/pnas.1600339113}{Proc. Nat. Acad. Sci. U.S.A. {\bf 113}, 5498--5502 (2016)}.

%\bibitem{Giomi:2013a}
%L. Giomi, N. Hawley-Weld, and L. Mahadevan,
%Swarming, swirling and stasis in sequestered Bristle-Bots,
%\href{http://dx.doi.org/10.1098/rspa.2012.0637}{Proc. R. Soc. A {\bf 469}, 20120637 (2013)}.

%\bibitem{Shraiman:2005}
%B. I. Shraiman,
%Mechanical feedback as a possible regulator of tissue growth.
%\href{http://dx.doi.org/10.1073/pnas.0404782102}{Proc. Nat. Acad. Sci. U.S.A. {\bf 102}, 3318--3323 (2005)}.

%\bibitem{Montel:2011}
%F. Montel, M. Delarue, J. Elgeti, L. Malaquin, M. Basan, T. Risler, B. Cabane, D. Vignjevic, J. Prost, G. Cappello, J. F. Joanny,
%Stress clamp experiments on multicellular tumor spheroids.
%\href{http://dx.doi.org/10.1103/PhysRevLett.107.188102}{Phys. Rev. Lett. {\bf 107}, 188102 (2011)}.

%\bibitem{Kumar:2013}
%P. Kumar, A. Libchaber,
%Pressure and temperature dependence of growth and morphology of \textit{Escherichia Coli}: experiments and stochastic model.
%\href{http://dx.doi.org/10.1016/j.bpj.2013.06.029}{Biophys J. {\bf 105}, 783---793 (2013)}.

%\bibitem{Chaikin:1995}
%P. M. Chaikin, T. C. Lubensky,
%\emph{Principles of condensed matter physics}
%(Cambridge University Press, Cambridge, England, 1995).

%\bibitem{Bowick:2008}
%M. J. Bowick, L. Giomi, H. Shin, C. K. Thomas,
%Bubble-raft model for a paraboloidal crystal.
%\href{http://dx.doi.org/10.1103/PhysRevE.77.021602}{Phys. Rev. E {\bf 77}, 021602 (2008)}.

%\bibitem{Duvernoy:2018}
%M. C. Duvernoy, T. Mora, M. Ardré, V. Croquette, D. Bensimon, C. Quilliet, J. M. Ghigo, M. Balland, C. Beloin, S. Lecuyer, N. Desprat,
%Asymmetric adhesion of rod-shaped bacteria controls microcolony morphogenesis.
%\href{https://www.nature.com/articles/s41467-018-03446-y}{Nat. Commun. {\bf 9}, 1120 (2018)}.

%\bibitem{Landau:1986}
%L. D. Landau and E. M. Lifshitz,
%Theory of Elasticity, 3rd ed.
%(Butterworth-Heinemann, Oxford, 1986).

%\bibitem{Hatwalne:2004}
%Y. Hatwalne, S. Ramaswamy, M. Rao, R. A. Simha,
%Rheology of active-particle suspensions.
%\href{http://dx.doi.org/10.1103/PhysRevLett.92.118101}{Phys. Rev. Lett. {\bf 92}, 118101 (2004)}.

%\bibitem{Ahmadi:2005}
%A. Ahmadi, T. B. Liverpool, M. C. Marchetti,
%Nematic and polar order in active filament solutions.
%\href{https://doi.org/10.1103/PhysRevE.72.060901}{Phys. Rev. E {\bf 72}, 060901(R)}.

%\bibitem{Ahmadi:2006}
%A. Ahmadi, M. C. Marchetti, T. B. Liverpool,
%Hydrodynamics of isotropic and liquid crystalline active polymer solutions.
%\href{http://dx.doi.org/10.1103/PhysRevE.74.061913}{Phys. Rev. E {\bf 74}, 061913 (2006)}.

%\bibitem{Liverpool:2007}
%T. B. Liverpool, M. C. Marchetti,
%Hydrodynamics and rheology of active polar filaments.
%in {\em Cell Motility}, edited by P. Lenz (Springer, New York, 2007).

%\bibitem{Giomi:2013b}
%L. Giomi, M. J. Bowick, X. Ma, and M. C. Marchetti,
%Defect annihilation and proliferation in active nematics.
%\href{http://doi.org/10.1103/PhysRevLett.110.228101}{Phys. Rev. Lett. {\bf 110}, 228101 (2013)}.

%\bibitem{Giomi:2014}
%L. Giomi, M. J. Bowick, P. Mishra, R. Sknepnek, and M. C. Marchetti,
%Defect dynamics in active nematics.
%\href{http://dx.doi.org/10.1098/rsta.2013.0365}{Phil. Trans. R. Soc. A {\bf 372}, 20130365 (2014)}.

%\bibitem{Lau:2009}
%A. W. C. Lau, T. C. Lubensky,
%Fluctuating hydrodynamics and microrheology of a dilute suspension of swimming bacteria.
%\href{http://dx.doi.org/10.1103/PhysRevE.80.011917} {Phys. Rev. E {\bf 80}, 011917 (2009)}.

%\bibitem{Doostmohammadi:2015}
%A. Doostmohammadi, S. P. Thampi, T. B. Saw, C. T. Lim, B. Ladoux, J. M. Yeomans,
%Cell division: a source of active stress in cellular monolayers.
%\href{http://dx.doi.org/10.1039/C5SM01382H}{Soft Matter {\bf 11}, 7328--7336 (2015)}.

%\bibitem{Marschall:2018}
%Theodore Marschall and S. Teitel,
%{\em Compression-Driven Jamming of Athermal Frictionless Spherocylinders in Two Dimensions},
%\href{https://doi.org/10.1103/physreve.97.012905}{Phys. Rev. E {\bf 97}(1), 012905 (2018)}.

%\bibitem{Chen:2017}
%S. Chen, P. Gao, T. Gao,
%Dynamics and structure of an apolar active suspension in an annulus.
%\href{https://doi.org/10.1017/jfm.2017.759}{J. Flui. Mech. {\bf 835}, 393--405 (2017)}.

%\bibitem{Ockendon:2003}
%J. Ockendon, S. Howison, A. Lacey, and A. Movchan, 
%{\em Applied Partial Differential Equations}
%(Oxford University Press, Oxford, 2003).

%\bibitem{Lianou:2013}
%K. P. Koutsoumanis, A. Lianou,
%Stochasticity in colonial growth dynamics of individual bacterial cells.
%\href{http://aem.asm.org/content/79/7/2294.short}{Appl. Environ. Microbiol. {\bf 79}, 2294--2301 (2013)}.

%\bibitem{Smith:2017}
%W. P. J. Smith, Y. Davit, J. M. Osbornec, W. Kim, K. R. Fosterd, J. M. Pitt-Francis,
%Cell morphology drives spatial patterning in microbial communities.
%\href{www.pnas.org/cgi/doi/10.1073/pnas.1613007114}{Proc. Nat. Acad. Sci. U. S. A. {\bf 114}, E280 (2017)}.

\end{thebibliography}
\end{document}

\title{Supplemental Material: Confinement-induced self-organization in growing bacterial colonies}

\author{Zhihong You}
\affiliation{Department of Physics, University of California Santa Barbara, Santa Barbara, CA 93106, USA}
\affiliation{Instituut-Lorentz, Universiteit Leiden, P.O. Box 9506, 2300 RA Leiden, The Netherlands}
\author{Daniel J. G. Pearce}
\affiliation{Department of Theoretical Physics, Universit\'e de Gen\`eve, 1205 Gen\`eve, Switzerland}
\author{Luca Giomi}
\email[Corresponding author:\ ]{giomi@lorentz.leidenuniv.nl}
\affiliation{Instituut-Lorentz, Universiteit Leiden, P.O. Box 9506, 2300 RA Leiden, The Netherlands}

\maketitle

\section*{Continuum Model}

Our continuum model is based on Eqs. (8) in the main text, describing the dynamics of the mass density field $\rho$, the momentum density $\rho\bm{v}$, the nematic tensor $\bm{Q}$ and the stress tensor $\bm{\sigma}$. These equations are integrated in a square $L \times L$ grid of lattice spacing $\Delta/L=5\times 10^{-3}$. Time integration is performed via a 4th order Runge-Kutta algorithm and interrupted after $10^7$ time steps, after the system has reached a steady state. The integration domain is endowed with stress-free absorbing boundaries on the $x$-direction and periodic boundaries on the $y$-direction. This implies: 
\begin{subequations}
\begin{align}
\partial_{x}
\left(
\begin{array}{c}
\rho\\[5pt]
v_{x}\\[5pt]
v_{y}\\[5pt]
Q_{xx}\\[5pt]
Q_{xy}
\end{array}
\right)_{x=\pm \frac{L}{2}}
&= \left(
\begin{array}{c}
0\\[5pt]
0\\[5pt]
0\\[5pt]
0\\[5pt]
0
\end{array}
\right)\;,\\[5pt]
\left(
\begin{array}{c}
\tilde{\sigma}_{xx} \\[5pt]
\tilde{\sigma}_{xy} \\
\end{array}
\right)_{x=\pm\frac{L}{2}}
&= \left(
\begin{array}{c}
0 \\[5pt]
0 \\
\end{array}
\right)\;.
\end{align}
\end{subequations}
Unless otherwise stated, the numerical values of material parameters appearing in Eqs. (8) are given in Table \ref{tab:parameters}. To render the parameters dimensionless, we rescale length by $L$, time by the characteristic relaxation time of the nematic phase $\tau_{n} = \gamma L^{2}/K$ and stress by $\sigma=K/L^2$. All other other quantities are rescaled accordingly.

\begin{table}[b]
	\begin{ruledtabular}
	\begin{tabular}{c c c}
		{\bf Symbol} & {\bf Name} & {\bf Value} \\ 
		\hline
		$k_g$ 		& Growth rate 							& $0.8$					\\[5pt] 
		$\eta$ 		& Shear viscosity 						& $0.25$				\\[5pt] 
		$\xi$ 		& Drag per unit mass 					& $400$					\\[5pt]
		$\alpha_0$ 	& Active stress 						& $-400$				\\[5pt] 
		$\tau$ 		& Viscoelastic crossover time 			& $1.25 \times 10^{-3}$	\\[5pt]
		$\kappa$ 	& Strain-orientation coupling 	   		& $2$					\\[5pt]
		$P_0$ 		& Pressure								& $800$					\\[5pt]
		$\rho_c$ 	& Isotropic-nematic transition density	& $0.05$				\\[5pt]
		$\rho_0$ 	& Stress propagation density 		 	& $0.5$ 
	\end{tabular}
	\end{ruledtabular}
\caption{\label{tab:parameters}Numerical values of the material parameters appearing in Eqs. (5) expressed in the dimensionless units given in the text.}
\end{table}

\section*{Role of channel width in unconfined colony}

\begin{figure*}
\centering
\includegraphics[width=1.0\textwidth]{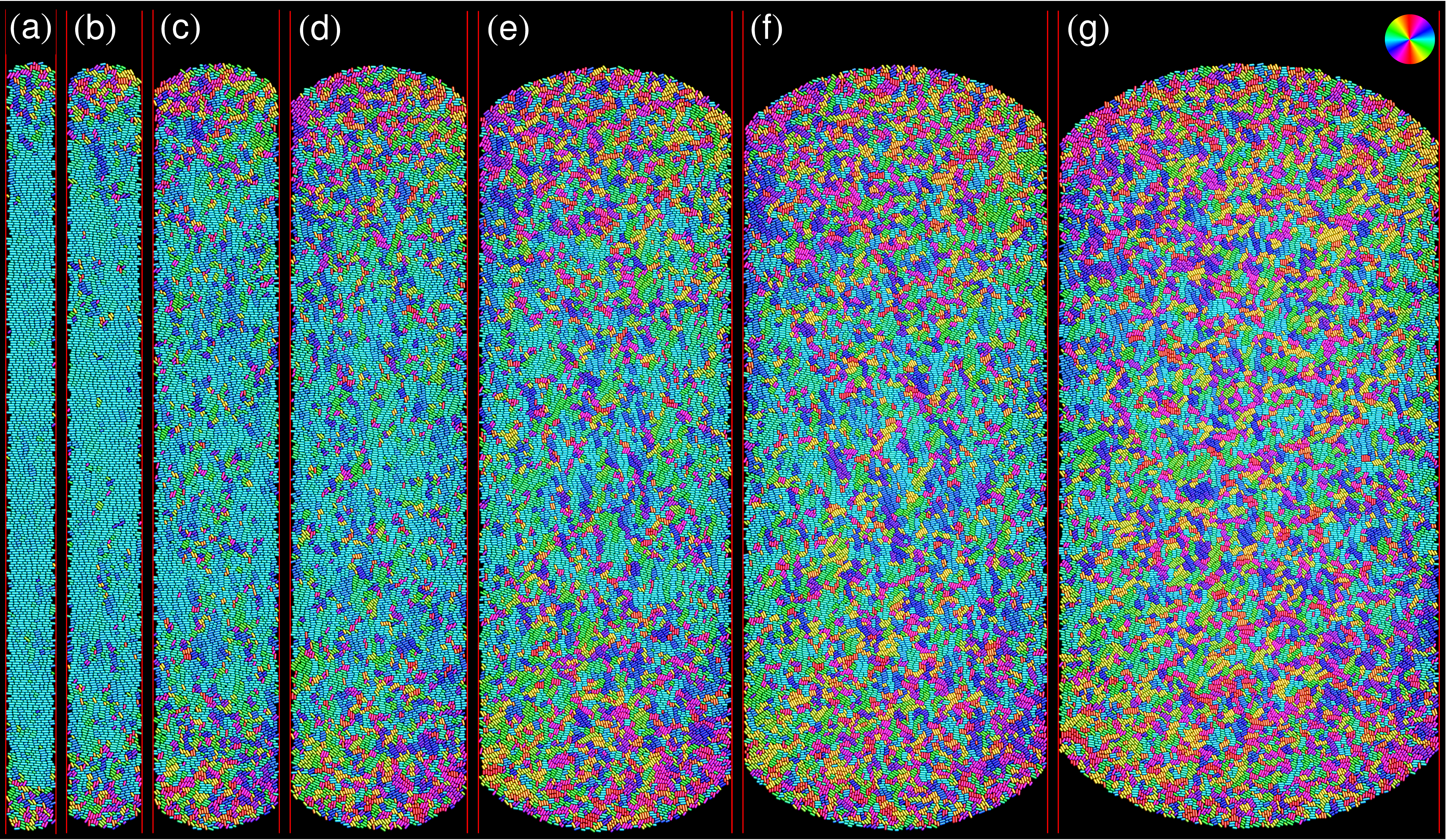}
\caption{\label{fig:ss-infi-lxs} \textbf{Effect of the channel width.} Snapshots of the unconfined colony with different $L_{x}$ values: (a) 20 $\mu$m, (b) 40 $\mu$m, (c) 50 $\mu$m, (d) 70 $\mu$m, (e) 100 $\mu$m, (f) 120 $\mu$m, (g) 150 $\mu$m. The heights of all panels are $300 \mu$m. Cells are colored coded by their orientations, as indicated by the color wheel in panel (g).}
\end{figure*}

Figs. \ref{fig:ss-infi-lxs}a--\ref{fig:ss-infi-lxs}g show examples of the long time configuration attained by unconfined colonies of various width $L_{x}$. For all $L_{x}$ values, the colonies' bulk eventually approaches a steady state where the alignment parameter $\Phi$ remains constant in time. In narrow channels (e.g. Figs. \ref{fig:ss-infi-lxs}a--\ref{fig:ss-infi-lxs}b), $\Phi=1$ and cells are perfect aligned along the horizontal direction. Increasing the width $L_{x}$ decreases the anisotropy in the normal stresses, resulting into less ordered bulk configurations delimited by thicker and more disordered caps. 

\begin{figure*}
	\centering
	\includegraphics[width=0.65\textwidth]{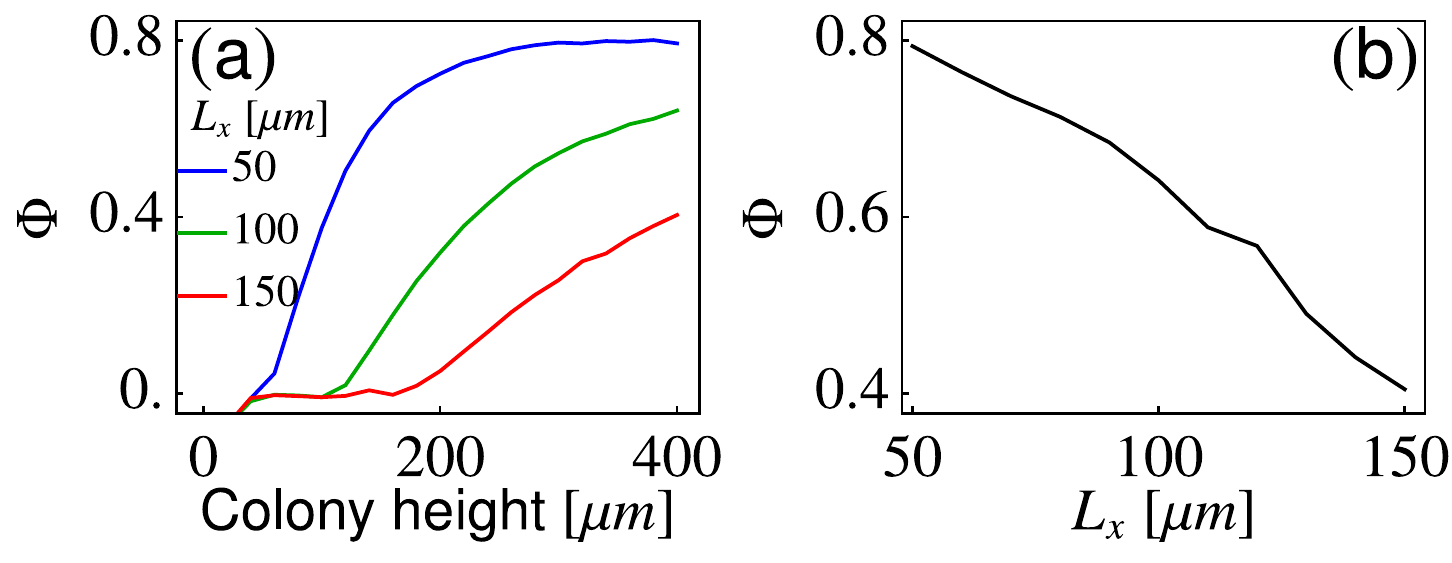}
	\caption{\label{fig:ALxLy} \textbf{Alignment parameter in region $\mathcal{R}_{0}$ (as defined in the main text) at the center of an expanding, unconfined colony.} (a) The alignment increases as a function of colony height. This increase is faster for short channels. (b) Alignment parameter in $\mathcal{R}_{0}$ and at a colony height of $400\,\mu$m, as a function of channel width $L_x$.}
\end{figure*}

Fig.~\ref{fig:ALxLy}a shows that the alignment parameter increases faster for channels of smaller width. This is because for a short channel the absorbing boundaries efficiently relieve the perpendicular stress $\sigma_{xx}$. Therefore the colony does not need to expand as far in the $y$ direction before there is a sufficient stress anisotropy to drive bacterial alignment. Fig.~\ref{fig:ALxLy}b shows this trend for a colony that has reached a height of $400\,\mu$m.

\section*{Role of cell aspect ratio}
\begin{figure*}
\centering
\includegraphics[width=1.0\textwidth]{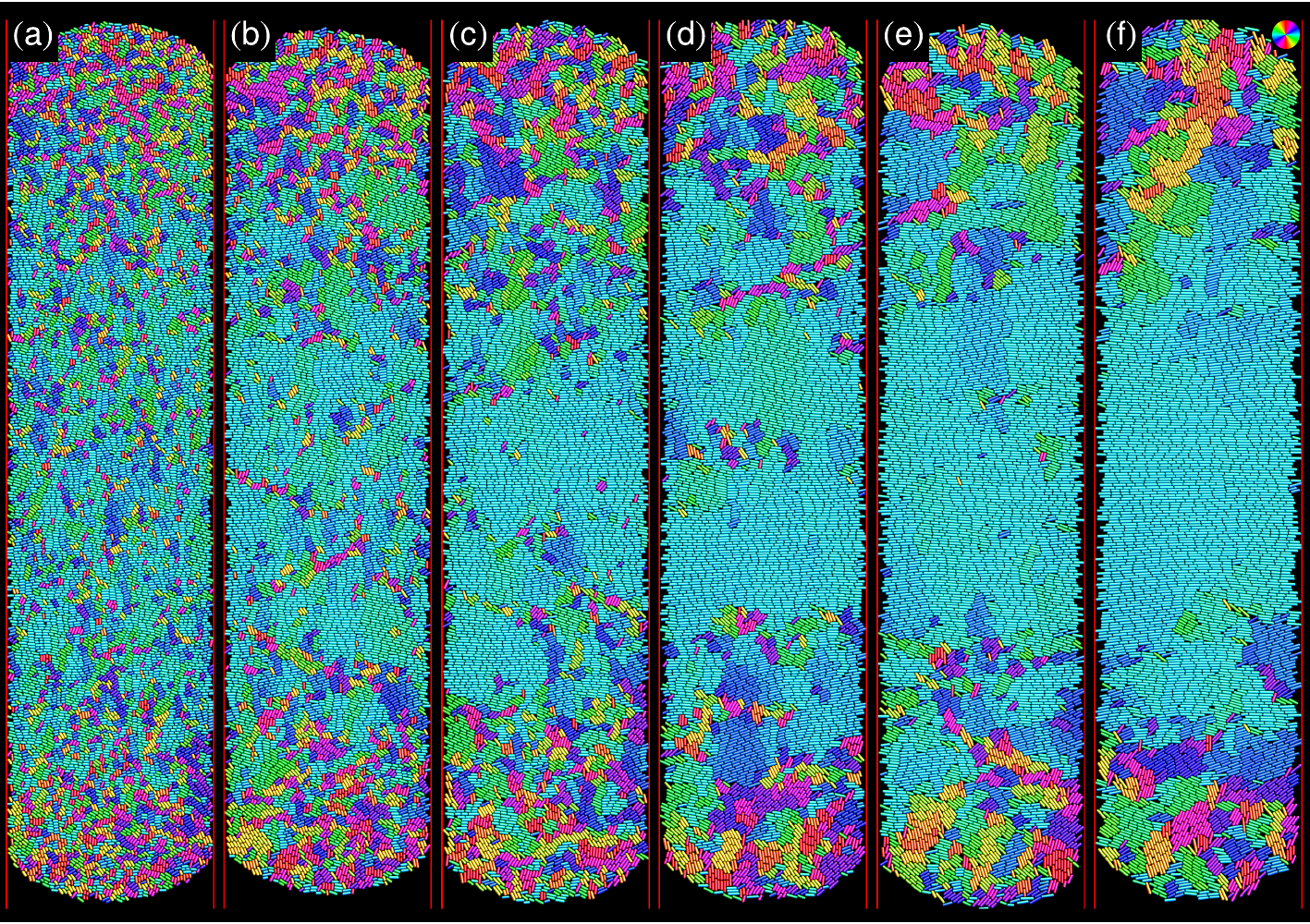}
\caption{\label{fig:ss-infi-lds} \textbf{Effect of the cells aspect ratio.} Snapshots of the unconfined colony with different $l_{d}/d_{0}$ values: (a) 2, (b) 2.5, (c) 3, (d) 3.5, (e) 4, (f) 4.5. The channel dimension is $70 \mu{\rm m} \times 300 \mu{\rm m}$. Cells are colored-coded by their orientations, as indicated by the color wheel in panel (f).}
\end{figure*}

Figs. \ref{fig:ss-infi-lds}a--\ref{fig:ss-infi-lds}f illustrate the effects of cell aspect ratio $l_{d}/d_{0}$. As explained in greater detail in Ref. [14], increasing the aspect ration results in an increase of the colony orientational stiffness, i.e. the constant $K$ in Eqs. (9) and (10). This, in turn, improves the performance of the aligning mechanism by rendering spatial variation of the nematic director energetically more costly. Furthermore, for a given stress anisotropy $\Delta\Sigma$, the larger the aspect ratio the higher the  torque experienced by the cells, thereby facilitating the alignment mechanism. Notably, for sufficiently large enough cell aspect ratio, a nearly perfect horizontal alignment can be found at the colony center, similar to those at narrow channels (e.g., Fig. \ref{fig:ss-infi-lxs}a).

\section*{Emergent flow}

\begin{figure*}
\centering
\includegraphics[width=1.0\textwidth]{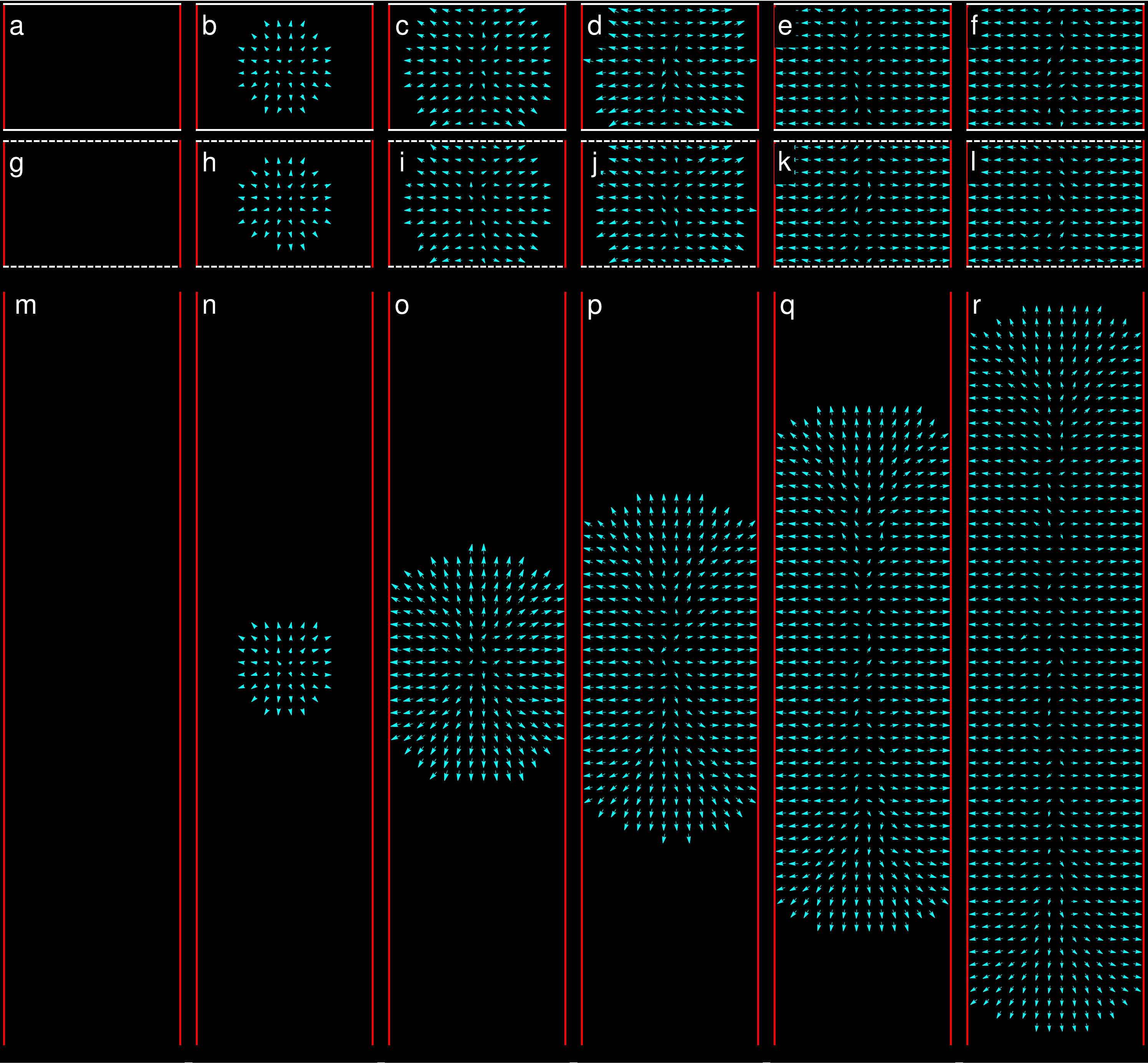}
\caption{\label{fig:velocity-grow} \textbf{Expansion flow in growing colonies.} Velocity field of the growing colonies shown in Fig. 1 of the main text. The boundary conditions are:  (a-f) rigid wall confinement, (g–l) periodic confinement and (m–r) no confinement in the vertical direction. In panels (a), (g), and (m), the cells are at rest since the colony contains only one cell. Each arrow shows the mean velocity of cells within a $5\,\mu{\rm m} \times 5\,\mu{\rm m}$ box centered at the arrow tail.}
\end{figure*}

Despite bacteria being non-motile, the interplay between cell growth and cell reorientation gives rise to interesting collective cell motion within the colony. Specifically, before coming to contact with the boundaries, freely expanding colonies are approximatively isotropic and the flow field resulting from cell division is radial on average (Figs. \ref{fig:velocity-grow}b, \ref{fig:velocity-grow}h, and \ref{fig:velocity-grow}n). As the colony reaches the boundaries, the vertical component of the velocity vanishes, whereas the horizontal component is enhanced by the cells' persistent growth and division (Figs. \ref{fig:velocity-grow}c-f, \ref{fig:velocity-grow}i-l and \ref{fig:velocity-grow}o-r). This process is faster in the presence of hard walls (Figs. \ref{fig:velocity-grow}a-f) or periodic boundaries (Figs. \ref{fig:velocity-grow}g-l) and more gradual in unconfined colonies (Figs. \ref{fig:velocity-grow}m-r). As detailed in the main text, the breaking of rotational symmetry in the flow leads to the emergence of globally anisotropic stress, hence further enhancing cell reorientation.